\documentclass{aa}  
\usepackage{graphicx}
\usepackage{txfonts}
\usepackage{natbib,twoopt}
\usepackage{color}
\usepackage{amsmath}
\usepackage{lscape}
\usepackage{pdflscape}
\usepackage{rotating}
\usepackage{subfiles}
\usepackage{stfloats}
\usepackage{epsfig} 

\usepackage{natbib}
\usepackage{txfonts}
\usepackage{balance}
\bibpunct{(}{)}{;}{a}{}{,}  

\begin{document} 

\def\sar{\ensuremath{\mathcal{\lambda_{\rm SAR}}}}
\newcommand{\dd}{\mathrm{d}}

   \title{AGN host galaxy mass function in COSMOS:}

   \subtitle{is AGN feedback responsible for the mass-quenching of galaxies?}

  \author{A. Bongiorno,\thanks{E-mail: \texttt{angela.bongiorno@oa-roma.inaf.it (OAR)}}
          \inst{1},
          A. Schulze,\inst{2}
          A. Merloni,\inst{3}
          G. Zamorani,\inst{4}
          O. Ilbert,\inst{5}
          F. La Franca, \inst{6}
          Y. Peng,\inst{7}
          E. Piconcelli,\inst{1}
          V. Mainieri, \inst{8}
           J. D. Silverman, \inst{2}
          M. Brusa,\inst{9,4}
          F. Fiore,\inst{1}
          M. Salvato, \inst{3}
          N. Scoville \inst{10}}

\institute{INAF-Osservatorio Astronomico di Roma, via Frascati 33, 00040 Monteporzio Catone, Italy;
\and
  Kavli Institute for the Physics and Mathematics of the Universe, Todai Institutes for Advanced Study, the University of Tokyo, Kashiwa, Japan 277-8583 (Kavli IPMU, WPI);
\and  
  Max-Planck-Institut fuer Extraterrestrische Physik (MPE), Postfach
1312, 85741 Garching, Germany;
\and
INAF-Osservatorio Astronomico di Bologna, via Ranzani 1, 40127
Bologna, Italy;
\and
Aix Marseille Universit\'e, CNRS, LAM (Laboratoire d'Astrophysique de Marseille) UMR 7326, 13388, Marseille,
France;
\and
Dipartimento di Matematica e Fisica, Universit\`a Roma Tre, via della Vasca Navale 84, 00146 Roma, Italy;
\and
Cavendish Laboratory, University of Cambridge, 19 J. J. Thomson
Ave., Cambridge CB3 0HE, UK;
\and 
European Southern Observatory, Karl-Schwarzschild-str. 2, 85748 Garching bei Muenchen, Germany;
\and
Dipartimento di Fisica e Astronomia, Universit\`a di Bologna, viale Berti Pichat 6/2, 40127 Bologna, Italy;
\and 
California Institute of Technology, MC 249-17, 1200 East
California Boulevard, Pasadena, CA 91125
 }

   \date{Received September 15, 1996; accepted March 16, 1997}


\authorrunning{A. Bongiorno et al.}
\titlerunning{AGN Host Galaxy Mass Function}

\abstract{We investigate the role of supermassive black holes in the global context of galaxy evolution by measuring the host galaxy stellar mass function (HGMF) and the specific accretion rate  i.e., \sar,  distribution function (SARDF) up to z$\sim$2.5 with $\sim$1000 X-ray selected AGN from XMM-COSMOS. Using a maximum likelihood approach, we jointly fit the stellar mass function and specific accretion rate distribution function, with the X-ray luminosity function as an additional constraint.
Our best fit model characterizes the SARDF as a double power-law with mass dependent but redshift independent break whose low \sar\  slope flattens with increasing redshift while the normalization increases. This implies that, for a given stellar mass, higher \sar\ objects have a peak in their space density at earlier epoch compared to the lower \sar\ ones, following and mimicking the well known AGN cosmic downsizing as observed in the AGN luminosity function.
The mass function of active galaxies is described by a Schechter function with a almost constant $\rm M^*_{\star}$ and a low mass slope $\alpha$ that flattens with redshift. Compared to the stellar mass function, we find that the HGMF has a similar shape and that, up to $\rm log(M_{\star}/M_{\odot})\sim$11.5 the ratio of AGN host galaxies to star forming galaxies is basically constant ($\sim$10\%). 
   Finally, the comparison of the AGN HGMF for different luminosity and specific accretion rate sub-classes with the phenomenological model prediction by Peng et al., 2010 for the ``transient'' population, i.e. galaxies in the process of being mass-quenched, reveals that low-luminosity AGN do not appear to be able to contribute significantly to the quenching and that at least at high masses, i.e.  $M_\star>10^{10.7} M_{\odot}$, feedback from luminous AGN ($\rm \log L_{bol}\gtrsim46$\,[erg/s]) may be responsible for the quenching of star formation in the host galaxy. }

\maketitle  
%

\section{Introduction}

Super-massive black hole (SMBH) growth, nuclear activity, and galaxy evolution, have been found to be closely related. In fact, over the last 15 years, the discovery of tight correlations between galaxies and their central nuclei properties \citep[see][and references therein]{Kormendy2013} as well as similar evolutionary trends between the growth histories of SMBHs and galaxies \citep[e.g.][]{Boyle1998, Marconi2004}, have established a new paradigm in which Active Galactic Nuclei (AGN) are key players in the process of galaxy formation and evolution. 
Several theoretical models \citep[e.g.][]{Somerville2001,Granato2004,Monaco2005,Springel2005,Croton2006,Hopkins2006a,Schawinski2006,Cen2011} have been developed to explain this co-evolution, and to find the mechanism responsible for the simultaneous fuelling of the central BH and the formation of new stars in the host galaxy, as well as the quasi-simultaneous shut-off of both processes. While the physical scales of interest (a few pc) cannot be directly resolved in these models and in current numerical simulations \citep[e.g.][]{Sijacki2015}, they usually propose the presence of an energetic AGN-driven feedback, i.e. a strong wind originated from the AGN that deposits the energy released by the accretion process within the host galaxy  \citep{Faucher2012}. This mechanism is able to link black hole growth and star formation and shut off both processes in a self-regulated manner. However, it is still unclear and observationally not proven, whether AGN driven feedback processes do indeed have an effect on the global properties of the galaxy population, in particular in suppressing the star formation (SF) in their host galaxies heating and/or pushing away the gas which is forming stars. 

Star formation quenching via some mechanism is required also to prevent the overgrowth of massive galaxies, hosted in the most massive dark-matter haloes \citep[e.g.][]{Read2005}.
Such a ``mass quenching'' mechanism, irrespective of  its physical origin, would suppress the growth of massive galaxies and explain the steep decline of the galaxy mass function above a given characteristic mass. While supernova feedback is not energetic enough in this mass regime, a central AGN would be an efficient mechanism. 

To investigate such a role for AGN, detailed studies on single objects have been performed to search for signatures of AGN feedback. Massive outflows on several kpc scales have been observed in a few cases \citep{Cano-Diaz2012,Feruglio2010,Cresci2015,Feruglio2015}, but up to now the evidence that such outflows are indeed responsible for suppressing star formation in the region of the outflow is circumstantial \citep{Cano-Diaz2012,Cresci2015,Cresci2015b}.
Further progress can be made through statistical studies of the properties of active galaxies (e.g., SFR) compared to normal galaxies. 
However, results have been often contradictory, i.e. some authors found that AGN  mainly lie above or on the Main Sequence (MS) of galaxies \citep{Santini2012,Mullaney2012}, while others \citep{Bongiorno2012,Mullaney2015} found the SFR of AGN hosts to be lower than the average MS galaxies, as expected by the models including AGN feedback. \citet{Bundy2008} compared the star formation quenching rate with the rate at which AGN activity is triggered in galaxies, and showed that these two quantities agree over a range of masses. They interpret this as a physical link between these two phenomena which however do not directly imply a causal link.

In fact, irrespective of AGN feedback, an essential pre-requisite to understand the role of black hole activity in galaxy evolution is to have a accurate and unbiased census of the AGN population an its relation to the properties of their host galaxies. The former is basically provided by the AGN luminosity function, which is now well established over a wide range of redshift and luminosity \citep{Ueda2014,Buchner2015,Aird2015,Miyaji2015,Silverman2008}. Deep X-ray surveys established a trend of AGN downsizing, i.e. the most luminous AGN have the peak in their space density at earlier times than lower luminosity AGN \citep{Ueda2003,Hasinger2005}, which is also seen in optical surveys \citep{Bongiorno2007,Croom2009}. This trend is similar to the downsizing in the galaxy population \citep{Cowie1996}, where the most massive galaxies build their mass at earlier times than lower mass galaxies. 

Linking black hole growth to their host galaxies, requires the study of their stellar mass function and/or the active fraction or duty cycle of AGN occurrence in galaxies of given stellar mass \citep[e.g.][]{Bundy2008,Xue2010,Georgakakis2011,Aird2012,Bongiorno2012,Lusso2012}. Most of these studies define AGN activity above a certain X-ray luminosity threshold and found the fraction of AGN at given $L_X$ to increase with stellar mass. However, this may lead to a biased view, since AGN at different masses cover different ranges of Eddington ratios for a given luminosity range and AGN have been found to show a wide distribution of Eddington ratios \citep[e.g.][]{Kauffmann2009,Schulze2010}. In fact, \citet{Aird2012} showed that the intrinsic distribution of specific accretion rates at $z<1$ follows a power law, whose shape does not evolve with redshift, independent of stellar mass. This result has been confirmed and extended out to $z<2.5$ by \citet{Bongiorno2012}.

In this work, we build upon the aforementioned studies of AGN hosts by establishing the bivariate distribution function of stellar mass and specific accretion rate for a hard X-ray selected AGN sample over the redshift range $0.3<z<2.5$. We use the derived AGN host galaxy stellar mass function to test the hypothesis of AGN feedback as driver of star formation quenching. In particular, we test whether the AGN population can be associated and/or be responsible for mass quenching using the model prediction from \citet{Peng2010} for the mass function of the `transient population'' (i.e. galaxies in the process of being mass-quenched).

The paper is organized as follows: In Sec.~\ref{sec:sample} we present the X-ray selected sample we are using. Sec~\ref{sec:HGMF} presents the method used to derive the specific accretion rate distribution function and the AGN host galaxy mass function (Sec.~\ref{sec:ML} and \ref{sec:vmax}) and their results (Sec.~\ref{sec:results}). In Sec~\ref{sec:quenching}, we address the question of the link between AGN and star formation quenching by comparing the AGN host galaxy mass function, computed for different subsamples, with the model prediction for quenching galaxies by \citet{Peng2010}.

Throughout this paper, a standard cosmology ($\Omega_m$=0.3, $\Omega_{\lambda}$=0.7 and H$_0$=70 km s$^{-1}$ Mpc$^{-1}$) has been assumed. The stellar masses are given in units of
solar masses for a Chabrier IMF \citep{Chabrier2003}.

\begin{figure}
\begin{center}
\includegraphics[width=0.4\textwidth]{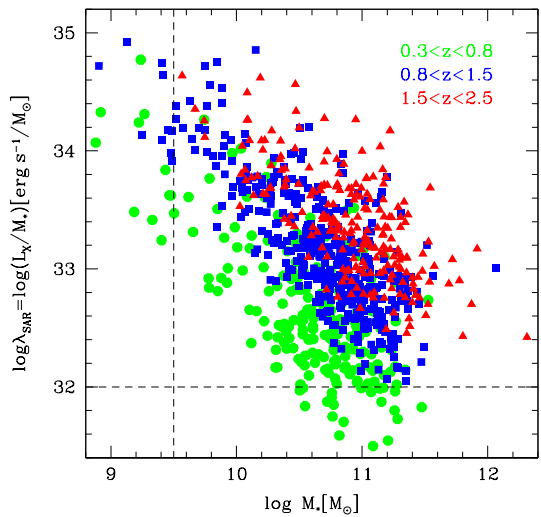}
\caption{Bivariate distribution for the analyzed hard X-ray selected sample in the M$_{\star}-\sar$ plane color-coded depending on the redshift bins. The horizontal and vertical dashed lines corresponds to the lower limit cuts applied in M$_{\star}$ and \sar.}
\label{fig:sample}
\end{center}
\end{figure}

\section{The Sample}
\label{sec:sample}

The AGN sample considered here has been extracted from the XMM-COSMOS point-like source
catalogue \citep{Hasinger2007,Cappelluti2009}  whose optical identifications and multiwavelength
properties have been presented by \citet{Brusa2010}. 
The catalog contains $\sim$1800 X-ray sources detected above flux limits of $\sim$5$\times$10$^{-16}$,
$\sim$3$\times$10$^{-15}$ and $\sim$7$\times$10$^{-15}$ erg cm$^{-2}$s$^{-1}$ in the [0.5-2] keV, [2-
10] keV and [5-10] keV bands, respectively.

Our analysis is based on objects that have been detected in the hard [2-10] keV band.
The restriction to a hard X-ray selected sample is chosen since the soft band can be affected by obscuration that can lead to a redshift-dependent incompleteness \citep[i.e. flux limited surveys pick up more obscured objects at higher redshift, see e.g.][]{Gilli2010}.  However, this band may still suffers from incompleteness due to  
heavily obscured and Compton Thick (CT, $\rm log(N_H)>{24}\,[cm^{-2}]$) AGN, whose detection probability is strongly reduced because the intrinsic emission can be significantly suppressed due to repeated Compton scattering and photoelectric absorption. 

Out of the full $\sim$1800 sources, we identify a final sample of 927 hard X-ray selected AGN in the redshift range 0.3$<$z$<$2.5.
All hard X-ray sources have accurate photometric redshifts  \citep{Salvato2011} while half (581/927) have secure spectroscopic redshifts.

\subsection{X-ray luminosities, host galaxy stellar masses and specific accretion rates}
\label{sec:sar}
Rest-frame, intrinsic X-ray [2-10] keV luminosities for the final sample have been derived from the observed hard X-ray flux. 
Following \citet{LaFranca2005}, we converted the observed [2-10] keV fluxes to the intrinsic [2-10] keV luminosities, for each AGN with a given measured N$_{\rm H}$, by applying a K-correction  computed by assuming an intrinsic X-ray spectrum with a photon index $\Gamma$=1.8, an exponential cut-off at E=200 keV and, a photoelectric absorption corresponding to the observed N$_{\rm H}$ column density.
The [2-10] keV luminosity is given by:
\begin{equation}
\rm L_{[2-10]keV}^{rf} = F_{[2-10]keV} 4 \pi D_L^2 K(z,N_H)
\end{equation}
where D$_{L}$ is the luminosity distance and $K(z,N_H)$ is the term which accounts for the K-correction and absorption correction. 
The absorbing column density N$_{\rm H}$ for our sample has been derived as in \citet{Merloni2014}. For the brightest sources (above 200~pn counts in the 0.5-10~keV band of \textit{XMM-Netwton}) N$_{\rm H}$ is obtained from the full spectral analysis of \citet{Mainieri2011}, which is available for 195/927 of the AGN. For the remaining sources, N$_{\rm H}$ is estimated in a statistical fashion, by assessing the value of the `observed' spectral slope from the hardness ratio and assessing the value of the 'observed' spectral slope drawn from a normal distribution with mean and dispersion of $\Gamma_{\rm int} = 1.8\pm0.2$. While this estimate shows a significant scatter, there are no apparent systematic biases, as demonstrated in \citet{Merloni2014}. Therefore these estimates can be robustly used for the statistical studies as performed in this paper. 
 
Host galaxy stellar masses have been derived in \citet{Bongiorno2012} using a two-component (AGN and galaxy) SED fitting technique. We refer the reader to this paper for a detailed description of the method.

Following \citet{Aird2012} and \citet{Bongiorno2012} we define \textit{``specific accretion rate''} \sar $\equiv  L_X/M_\star$ \citep[see also][]{Brusa2009,Georgakakis2014} as a directly measurable quantity which can be regarded as a proxy for the black hole growth rate relative to the stellar mass of the host galaxy, $\dot{M}_{\rm BH}/M_\star$, after taking into account the (luminosity dependent) bolometric correction \citep[e.g.][]{Marconi2004,Lusso2012} and a radiative efficiency factor. It is also related to the SMBH's Eddington ratio, $\lambda_{\rm{Edd}}=L_{\rm bol}/M_{\rm BH}$, applying the bolometric correction factor and the scaling relationship between black hole mass and host stellar mass. 
Assuming as an approximation a mean bolometric correction $k_{bol}$=25 \citep{Marconi2004,Lusso2012} and a constant host stellar to black hole mass ratio of 500 \citep{Marconi2003,Haring2004}, $\log\sar=34\,[erg/s/M_{\odot}]$ approximately corresponds to the Eddington limit, while $\log\sar=32\, [erg/s/M_{\odot}]$ would give 1\% of Eddington. 
The bivariate distribution M$_\star$ - \sar\ for the analyzed sample is shown in Fig. \ref{fig:sample} where different colors correspond to different redshift ranges as labeled.

For the determination of the mass function, we further restrict our sample in stellar mass $M_\star$ and specific accretion rate \sar, applying the following cuts: $\rm M_\star>10^{9.5}\ M_\odot$ and $\rm \sar>10^{32}\,erg/s/M_{\odot}$. 
The latter criterion is motivated by the requirement of having a clear cut in \sar\ above which we define the AGN as active (see below). The chosen minimum \sar\ value corresponds to the lowest observed value in our intermediate redshift bin and furthermore corresponds approximately to 1\% of Eddington which we chose in the following as our minimum threshold to define an active black hole, consistent with studies of type--1 AGN \citep{Schulze2015}. After applying these limits, our sample is reduced to 877 AGN with $0.3<z<2.5$.

 \begin{table*}
\caption{Best fit model parameters and their errors for the bivariate distribution function of stellar mass and SAR (eq. \ref{eq:psi_me}). The parameters denoted with an $^*$ are kept fixed during the fit.}
\label{tab:dfpara}
\centering
\begin{tabular}{c |c c  | c c c c c c | c c c}
\hline \hline 
&\multicolumn{2}{ |c |}{}&\multicolumn{6}{c |}{}&\multicolumn{3}{c }{}\\
&\multicolumn{2}{| c |}{$f_\star(M_\star,z)$ from eq. \ref{eq:schechter}} & \multicolumn{6}{c |}{$f_\sar(\sar)$ from eq. \ref{eq:dpl}} & \multicolumn{3}{c}{$f_z(z)$ from eq. \ref{eq:zevol}}\\ 
\hline
&\multicolumn{2}{| c |}{}&\multicolumn{6}{c |}{}&\multicolumn{3}{c }{}\\
$\log (\Psi^\ast)$  & $\log M_\star^*[M_{\odot}]$ & $\alpha$ & $\log\lambda^*_{\rm SAR,0}$ & $k_\lambda$ & $\log M_{\star,0}$ &$\gamma_{1,0}$ & $k_\gamma$  & $\gamma_2$ & $p_1$ & $p_2$ & $z_0$ \\ 
\hline 
&\multicolumn{2}{c |}{}&\multicolumn{6}{c |}{}&\multicolumn{3}{c }{}\\
-6.86 & 10.99 & 0.24 &  $33.8^*$ & -0.48 & $11.0^*$ & -1.01 & 0.58 & -3.72 & 5.82 & 2.36 & $1.1^*$  \\
+/-0.01 & +/-0.03 &  $^{+0.07}_{-0.06}$ & -- & $^{+0.03}_{-0.03}$ & -- & $^{+0.02}_{-0.02}$ & $^{+0.02}_{-0.02}$ & $^{+0.09}_{-0.09}$ & $^{+0.12}_{-0.13}$ & $^{+0.08}_{-0.08}$ & --\\
\noalign{\smallskip}
\hline
\hline
\end{tabular}
\end{table*}

\begin{figure*}
\begin{center}
\includegraphics[width=0.8\textwidth]{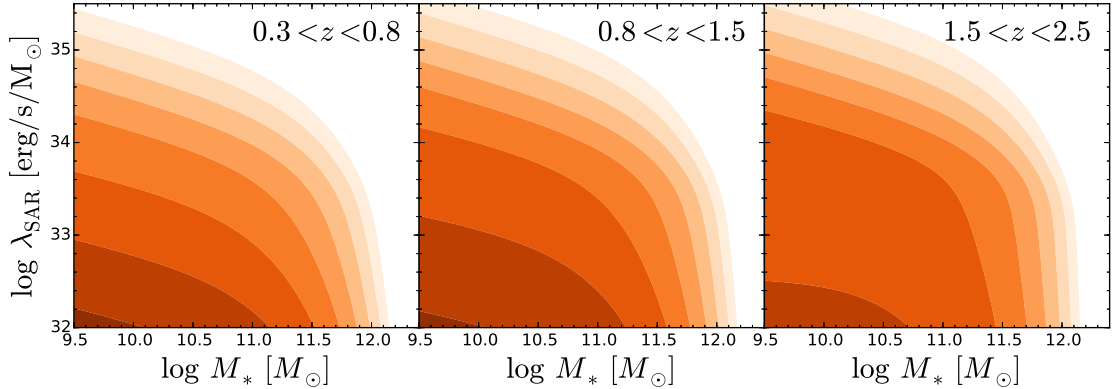}
\caption{Bivariate distribution function $\Psi(M_\star,\sar,z)$, for our best fitting parametric model,  derived through the Maximum Likelihood method, in three redshifts bins. The orange contours indicate  lines of constant space density, from $10^{-10}$ to $10^{-3}$ Mpc$^{-3}$), separated by a factor of 10 each.}
\label{fig:bivariate}
\end{center}
\end{figure*}

\section{The AGN Host Galaxy Mass Function and specific accretion rate distribution function}
\label{sec:HGMF}

In order to derive the AGN host galaxy mass function (HGMF) and the specific accretion rate distribution function (SARDF), we have to account for various selection effects in our flux-limited AGN sample. This requires a careful assessment of the incompleteness function.

In fact, completeness in L$_{\rm X}$ does not directly ensure completeness in $\rm M_\star$. 
As previously reported, AGN show a wide range of Eddington ratios \citep{Kauffmann2009,Schulze2010}, and thus also a wide range of $L_X/M_\star$ (\sar), with a distribution falling below the corresponding Eddington limit approximately following a power-law distribution \citep{Aird2012,Bongiorno2012}.

A luminosity complete AGN sample will be biased towards high mass BHs and high galaxy mass i.e. since an AGN with low Eddington ratio will be included in the sample only if its M$_{\rm BH}$ is high enough to be above the given luminosity (L$_X$) limit, given the relation between M$_{BH}$-M$_{\star}$, a bias towards high-mass black holes induces a bias toward high-mass galaxies.
This effect has to be carefully taken into account when building a galaxy mass complete sample starting from an X-ray flux-limited AGN sample.

\subsection{Incompleteness function}
\label{sec:inc_func}

Our corrections for incompleteness account for three effects: (1) the X-ray sensitivity function; (2) the absorption correction $ f(N_{H}\mid{L_{X}, z})$; and (3) the stellar mass completeness down to our threshold in units of specific accretion rate $\log\sar=32\, [erg/s/M_{\odot}]$.

The first selection effect to consider is the position dependent X-ray flux limit based on the sensitivity map computed by \citet{Cappelluti2009}. The absorption correction accounts for the sources which have been missed in the sample due to their high column density $N_H$. For this correction we use the $N_{H}$ distribution as a function of $z$ and $L_X$ published by \citet{Ueda2014} based on several  X-ray AGN surveys (see their eq. (5) and (6)). We integrate over the $N_{H}$ distribution between $20<\log N_H<24$, i.e. we do not include Compton thick AGN in our HGMF determination. The fraction of CT AGN is still uncertain and the $N_{H}$ distribution above $\log N_H=24$ is poorly known \citep{Ueda2014,Buchner2015,Aird2015}. The contribution of CT AGN to the AGN space density is expected to lie between $\sim10-40$\% \citep{Gilli2007,Treister2009,Vignali2014,Buchner2015,Lansbury2015}. These two corrections applied to the flux limited sample result in a luminosity complete sample.

As described above, we additionally suffer from significant incompleteness due to the fact that a broad range of $\rm M_\star$ can be associated to a given luminosity $\rm L_X$. To account for this effect in the HGMF, we need to include an additional term to the incompleteness function based on the distribution of \sar. Using this distribution function, we correct for incompleteness down to a fixed threshold in \sar, which we set at $\log\sar=32\, [erg/s/M_{\odot}]$. The HGMF is therefore defined as the mass function of all AGN above this \sar\ threshold.
The most rigorous and self consistent approach to do this is by determining the HGMF and the SARDF simultaneously, e.g. via the maximum likelihood method described in the next section.

\begin{figure*}
\begin{center}
\includegraphics[width=0.8\textwidth]{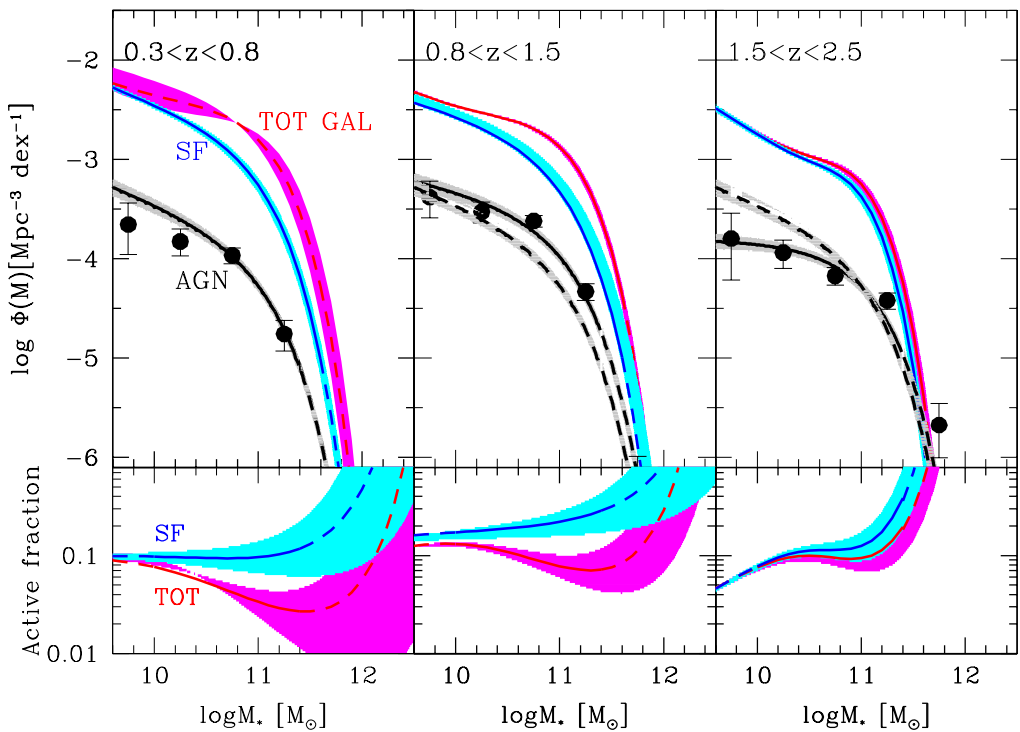}
\caption{\textit{Upper Panels:} Total AGN HGMF with the associated errors in three redshift bins derived through the  Maximum Likelihood (black line and grey shaded area) and the V$_{\rm max}$ (data points) methods. In each bin the lowest z fit is reported for reference with a dashed line. 
The red and the blue lines are the total and the star forming galaxy stellar mass functions with the associated errors are shown in magenta and cyan shaded areas) from \citet{Ilbert2013}.
\textit{Lower Panels:} Ratio of AGN host galaxies to the total (red line) and the star forming (blue line) galaxy population as a function of stellar mass in the same redshift bins.}
\label{fig:HGMF_ML_Vmax}
\end{center}
\end{figure*}

\subsection{Maximum Likelihood method}
\label{sec:ML}
 
We  here present the methodology of determining the SARDF and the HGMF simultaneously as a bivariate distribution function of stellar mass and specific accretion rate, i.e. $\Psi(M_\star,\sar,z)$, where $\Psi(M_\star,\sar,z) \dd \log M_\star \dd \log \sar$ gives the space density of AGN with stellar mass host galaxies between $\log M_\star$ and $\log M_\star+\dd\log M_\star$ and a specific accretion rate between $\log \sar$ and $\log \sar+\dd\log \sar$ at the redshift $z$. The HGMF, SARDF and the X-ray AGN LF (XLF) can be derived as different marginalizations over this bivariate distribution function.
We use the maximum likelihood method developed by \citet{Schulze2010} and extended by \citet{Schulze2015} to compute $\Psi(M_\star,\sar,z)$. While these works focused on the joint determination of the active black hole mass function and the Eddington ratio distribution function (using type 1 AGN), the method is implemented here for the joint determination of the HGMF and SARDF.

The technique minimizes the likelihood function $S=-2 \sum \ln p_i$, where the probability distribution $p_i$ for each object is given by:
\begin{eqnarray}
p_i(M_\star,\sar,N_H,z)  & = & \frac{1}{N} \Psi(M_\star,\sar,z)\, \mathcal{I}\left (M_\star, \sar, z, N_H\right ) \,  \nonumber\\ 
     & &        \times f\left (N_H\mid L_X, z\right ) \frac{\dd V}{\dd z}  \ ,
 \label{eq:pi}
\end{eqnarray}

where $\Psi(M_\star,\sar,z)$ is the bivariate distribution function of stellar mass and specific accretion rate that we want to derive, $\mathcal{I}\left (M_\star, \sar, z, N_H\right ) = \mathcal{I}(L_X,z,N_H)$ is the X-ray selection function given by the sensitivity map in the 2-10 keV band and $f\left (N_H\mid L_X, z\right )$ is the absorption distribution function, taken from \citet{Ueda2014}. We use the $N_H$ estimates presented in Sec.~\ref{sec:sar} to compute $L_X$ (and therefore \sar) and $f\left (N_H\mid L_X, z\right )$ for our sample.
The factor $N$ corresponds to the total number of objects in the sample predicted by the model and it is given by integrating over $M_\star$, \sar\ , N$_{\rm H}$, and $z$, i.e.
\begin{eqnarray}
&N = \iiiint \Psi(M_\star,\sar,z)\, \mathcal{I}\left (M_\star, \sar, z, N_H\right ) \,  \nonumber\\ 
      &         \times f\left (N_H\mid L_X, z\right ) \frac{\dd V}{\dd z}  \rm{dlog} N_H \rm{dlog}\sar \rm{dlog}M_\star \rm{d}z\ ,
 \label{eq:pi2}
\end{eqnarray}

\noindent
where we integrate over the $N_H$ distribution between $20<\log N_H<24$, while our integration ranges in $M_\star$, $\sar$ and $z$ are $9.5<\log M_\star<\infty$, $32<\log \sar<\infty$ and $0.3<z<2.5$, as discussed in Sec.~\ref{sec:sar}.

Our sample also contains 12 AGN without $M_\star$ measurements, due to poor quality photometry. However, we account for these sources using their luminosity and redshift information integrated over the entire mass range, i.e. 

\begin{equation}
p_j(L_X,N_H,z) = \int p_j(M_\star,\sar,N_H,z) \, \rm{dlog}M_\star \ .
\label{eq:pi_Lx}
\end{equation}

Our XMM-COSMOS based sample covers only a limited dynamical range in $L_X$, narrower than the full range over which the XLF is currently determined. This might lead to degenerate solutions for the bivariate distribution function, some of which may be inconsistent with the XLF. Ideally, we would like to construct the HGMF and SARDF including deeper and larger area surveys, but this is beyond the scope of the present work. To reduce this effect, we include as additional observational data the XLF. In this way, we ensure consistency with the XLF observations over its full observationally determined luminosity range. 
In particular, we use the binned XLF from \citet{Miyaji2015} and compute the $\chi^2$ value for the comparison with the XLF implied by the HGMF and SARDF. We then add this likelihood to that of the XMM-COSMOS sample.
The study by \citet{Miyaji2015} uses the same $N_H$ distribution as \citet{Ueda2014} for the determination of the XLF which we also employ here. Over our range in redshift and luminosity, the XLF by \citet{Miyaji2015} is consistent with other recent studies \citep{Ueda2014,Buchner2015,Aird2015}, thus our results are robust against the specific choice of XLF.

We caution that the faint end of the XLF is not directly constrained by our sample: the XLF will also include AGN below our threshold in $M_\star$ and \sar, which are not accounted for in our bivariate distribution function fit. This may lead to an overestimate of the space density at $\log L_X <43$~[erg/s].

 The total likelihood to minimize is given by:
 
 \begin{equation}
 S_{tot} = -2 \sum_{i=1}^{N_{M_\star}} \ln p_i(M_\star,\sar,N_H,z) - 2  \sum_{j=1}^{N_{L_X}} \ln p_j(L_X,N_H,z) + \chi^2(XLF) \ ,
 \label{eq:Stot}
\end{equation}
where $N_{M_\star}$ is the number of AGN with $M_\star$ measurements in our sample and $N_{L_X}$ is the number of AGN with only $L_X$ known.
The absolute normalization of the bivariate distribution function is then determined by scaling to the total observed number of objects in the sample.

Following \citet{Aird2012} and \citet{Bongiorno2012}, we first assume that the bivariate distribution function $\Psi(M_\star,\sar,z)$ is separable, i.e. the specific accretion rate distribution is mass independent and vice versa. Under this assumption, the bivariate distribution function is given by: 
\begin{equation}
\Psi(M_\star,\sar,z) = \Psi^* \,f_\sar(\sar,z)\,f_\star(M_\star,z)\,f_z(z) \ , \label{eq:psi_me1}
\end{equation}
where $\Psi^*$ is the normalization of the bivariate distribution function, $f_\sar(\sar,z)$ is the SAR-term, $f_\star(M_\star,z)$ is the M$_\star$-term and $f_z(z)$ is a redshift evolution term. 

However, for the SAR-term, we also tested a mass-dependent model and found this model to provide a better description of our data (see Appendix \ref{app:a} for more details). The bivariate distribution function is therefore written as:

\begin{equation}
\Psi(M_\star,\sar,z) = \Psi^* \,f_\sar(\sar,M_\star,z)\,f_\star(M_\star,z)\,f_z(z) \ , \label{eq:psi_me}
\end{equation}
\noindent
where $f_\sar(\sar,M_{\star},z)$ contains now also a dependence on the mass. We use this more general parametrization as our default model. We want to point out that the SAR-term $f_\sar$ and the M$_{\star}$-term $f_\star$ are not equal to the SARDF and HGMF.

The HGMF and the SARDF are calculated by integrating $\Psi(M_\star,\sar,z)$  over \sar\ and over $M_{\star}$, respectively. To be specific: 

\begin{eqnarray}
\Phi_\star(M_\star,z) = &\hspace{-1.0cm} \frac{dN}{dV\,dlogM}=\int_{32}^\infty \Psi(M_\star,\sar,z)\, \rm{dlog}\sar \nonumber\\
&\hspace{-1.2cm} = \int_{32}^\infty \Psi^* \,f_\sar(\sar,M_\star,z)\,f_\star(M_\star,z)\,f_z(z) \, \rm{dlog}\sar \
\label{eq:hgmf}
\end{eqnarray} 
and 
\begin{eqnarray}
\Phi_\sar(\sar,z) = &\hspace{-1.0cm} \frac{dN}{dV\,dlog\lambda}=\int_{9.5}^\infty \Psi(M_{\star},\sar,z)\, \rm{dlog}M_{\star} \nonumber\\
&\hspace{-1.4cm} = \int_{9.5}^\infty \Psi^* \,f_\sar(\sar,M_\star,z)\,f_\star(M_\star,z)\,f_z(z) \, \rm{dlog}M_{\star} \
\label{eq:sardf}
\end{eqnarray} 

In case of separable SAR- and M$_\star$-terms, as in Eq.~\ref{eq:psi_me1}, the SARDF (HGMF) has the same shape as $f_\sar$ ($f_\star$) and only the absolute normalization is determined by the marginalisation. However, in the more general case of Eq.~\ref{eq:psi_me}, this is not necessarily the case, which is why the HGMF and SARDF cannot then be explicitly expressed as analytic functions.



We here consider the following parametric models for the individual terms: the M$_\star$-term is modeled using a Schechter function:
\begin{equation}
f_\star(M_\star,z) = \left( \frac{M_\star}{M_\star^*} \right)^{\alpha} e^{\left( - \frac{M_\star}{M_\star^*} \ \right)}.  \  
\label{eq:schechter}
\end{equation} 
While a model with a low mass slope $\alpha$ evolving with redshift has been included, we find that the best fit parameters are indeed consistent with no z-evolution in $\alpha$.

The SAR-term is instead described by a double power law:
\begin{equation}
f_\sar(\sar,M_{\star},z) = \frac{1}{\left(\frac{\sar}{\lambda^*_{\rm SAR}(M_{\star})}\right)^{-\gamma_1(z)} + \left(\frac{\sar}{ \lambda^*_{\rm SAR}(M_{\star})}\right)^{-\gamma_2} } \ . 
\label{eq:dpl}
\end{equation}
where the low \sar\, slope $\gamma_1(z) = \gamma_{1,0} + k_\gamma (z-z_0)$, with $z_0$ set at $1.1$, and the break $\log \lambda^*_{\rm SAR}(M_{\star})=\log \lambda^*_{\rm SAR,0}+k_{\lambda} (\log M_\star-\log M_{\star,0})$ with $\log M_{\star,0}=11$.  \\
The assumption of a double power law for $\rm f_\sar$, allows to recover the double power law shape of the XLF with a Schechter function HGMF, as demonstrated by \citet{Aird2013}. We fixed the break value to $\rm log\lambda^*_{\rm SAR,0}=33.8\, [erg/s/M_{\odot}]$ to limit the number of free parameters.
This value is close to the implied Eddington limit, consistent with the approach in the study of \citet{Aird2013}, and with the tentative evidence for such a break first reported in \citet{Bongiorno2012}. 

Finally, we parameterize the redshift evolution of the normalization of the space density as:
\begin{equation}
f_z(z) = \begin{cases}
 (1+z)^{p_1} &  z\leq z_0 \\
 (1+z_0)^{p_1} \left( \frac{1+z}{1+z_0} \right)^{p_2}&  z>z_0 \\
 \end{cases}
 \label{eq:zevol}
\end{equation}
where we fixed $z_0=1.1$, motivated by the break redshift used in the LDDE model in the XLF from \citet{Miyaji2015} and approximately corresponding to the central redshift in our sample.

The best fit bivariate distribution function $\Psi(M_\star,\sar,z)$ is shown in Fig.~\ref{fig:bivariate} while the best fitting parameters and their errors are given in Table~\ref{tab:dfpara}. 
We computed the uncertainties of each parameter using a Markov chain Monte Carlo (MCMC) sampling of the likelihood function space, using \textit{emcee} \citep{Foreman2013}, a Python implementation of an Affine Invariant MCMC Ensemble sampler as presented by \citet{Goodman2010}. We used uniform priors for our free parameters and initilised the MCMC "walkers" around the best fit maximum likelihood solution. The quoted uncertainties represent the 16 and 84\%-tile of the parameter distribution, marginalized over all other parameters apart from $\Psi^*$. The latter is not determined by the Maximum likelihood fit and their error is given by $1/\sqrt{N_{\rm tot}}$.
 
As mentioned above, our best fit HGMF and SARDF given by Eq.~\ref{eq:hgmf} and Eq.~\ref{eq:sardf} cannot be expressed as simple analytic functions, due to the entanglement of $M_\star$ and \sar\ in the SARDF term. For a better quantitative representation of the redshift evolution of HGMF and SARDF and for illustrative purposes, we provide an analytic approximation of the two distribution functions, evaluated at the center of our three redshift bins. For this, at each redshift, we performed a least-squares fit to the HGMF (computed via Eq.~\ref{eq:hgmf}) with a standard Schechter function with normalisation $\Phi^\ast_{M}$, break $M_\star^*$ and low mass slope $\alpha$, and the SARDF (computed via Eq.~\ref{eq:sardf}) with a double power law with normalisation $\Phi^\ast_\lambda$, break $\lambda^*_{\rm SAR}$ and slopes $\gamma_{1}$, $\gamma_2$. We provide the best fit parameters in Tab.~\ref{tab:mf} and \ref{tab:sarf}.
 

\begin{table}
\caption{Best fit model parameters for the AGN host galaxy mass Schechter function, computed in our 3 redshift bins.}
\centering
\begin{tabular}{c c c c}
\hline \hline 
$<z>$  & $\log (\Phi^\ast_{M})$ & $\rm log M_\star^*$ & $\alpha$ \\ 
\noalign{\smallskip} \hline \noalign{\smallskip}
0.55 & $-3.83^{+0.04}_{-0.05}$ & $10.99^{+0.03}_{-0.03}$ &  $-0.41^{+0.04}_{-0.04}$ \\ 
\noalign{\smallskip}  \noalign{\smallskip}
1.15 & $-3.54^{+0.04}_{-0.05}$ & $10.99^{+0.03}_{-0.03}$ & $-0.24^{+0.04}_{-0.04}$   \\ 
\noalign{\smallskip}  \noalign{\smallskip}
2.00 & $-3.84^{+0.04}_{-0.04}$ & $10.99^{+0.03}_{-0.03}$ & $-0.03^{+0.05}_{-0.05}$     \\ 
\noalign{\smallskip}  \noalign{\smallskip}
 \hline \hline
\end{tabular}
\label{tab:mf}
\end{table}

\begin{table}
\caption{Best fit model parameters of the AGN specific accretion rate double power-law function, computed in our 3 redshift bins.}
\centering
\begin{tabular}{c  c c c c c c  }
\hline \hline 
\noalign{\smallskip}  \noalign{\smallskip}
$<z>$  &  $\log \Phi^\ast_\lambda$ & $\gamma_{1}$  & $\gamma_2$ &  $\rm log \lambda^*_{\rm SAR}$  \\ 
\noalign{\smallskip} \hline \noalign{\smallskip}
0.55 & $-6.04^{+0.08}_{-0.08}$  & $-1.35^{+0.02}_{-0.02}$ & $-3.64^{+0.10}_{-0.11}$ &  $34.33^{+0.04}_{-0.04}$ \\ 
\noalign{\smallskip} \noalign{\smallskip}
1.15 & $-5.22^{+0.08}_{-0.09}$  & $-1.02^{+0.02}_{-0.02}$ & $-3.61^{+0.10}_{-0.10}$ &  $34.32^{+0.03}_{-0.03}$ \\ 
\noalign{\smallskip}  \noalign{\smallskip}
2.00 & $-4.85^{0.08}_{-0.09}$  & $-0.54^{+0.03}_{-0.03}$ & -3.58$^{+0.10}_{-0.10}$ &  $34.30^{+0.03}_{-0.03}$\\
\noalign{\smallskip} \noalign{\smallskip}
 \hline \hline
\end{tabular}
\label{tab:sarf}
\end{table}

\begin{figure*}
\begin{center}
\includegraphics[width=0.8\textwidth]{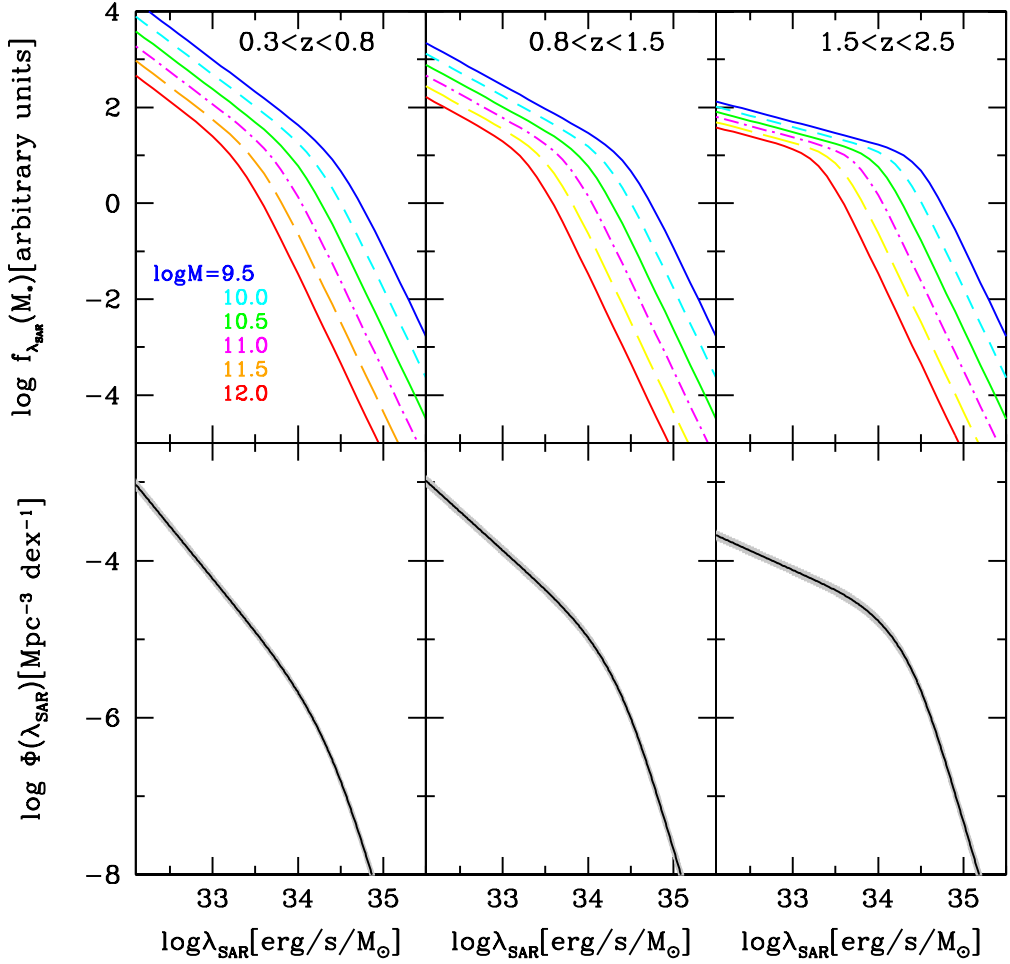}
\caption{\textit{Upper Panels:} The SAR-term split into three redshift bins and described as a double power-law with a mass dependent \sar$^*$ as in eq.~\ref{eq:dpl}. \textit{Lower Panels:} SARDF derived through the Maximum Likelihood method by integrating the bivariate distribution function over M$_{\star}$ (eq. \ref{eq:hgmf}). The shaded area includes the errors on the parameters.
}
\label{fig:Edd_ML}
\end{center}
\end{figure*}

\subsection{V$_{max}$ method}
\label{sec:vmax}
An additional consistency check can be obtained by computing the AGN host galaxy mass function using the V$_{max}$ method.
The V$_{\rm max}$ for each individual object is given by:

\begin{equation}
V_{\rm max}(M_\star) =\int_{z_{\rm min}}^{z_{\rm max}} \, \mathcal{A}(M_\star,z)\frac{dV}{dz}dz
\end{equation}

\noindent
where  $\mathcal{A}(M_\star,z)$ is the effective area as a function of $M_\star$ and $z$ given by the total survey area $\Omega$ times the incompleteness function. 
We emphasize here that the $V_{\rm max}(M_\star)$ values used are not identical to the $V_{\rm max}(L_{\rm X})$ values that would be used for the computation of the AGN luminosity function. This is because, as discussed above, we have to account in the incompleteness function also for the SARDF in addition to the sensitivity function and the absorption correction.
The incompleteness function thus includes three terms and can be written as:

\begin{eqnarray}
\mathcal{I}(M_\star,z)  =\int^{24}_{20}  \hspace{-0.1cm} \int^\infty_{\sar_\mathrm{min}} &\hspace{-0.6cm} \mathcal{I}\left (M_\star, \sar, z, N_H\right ) f\left (N_H\mid L_X, z\right ) \nonumber\\ 
                     &  \hspace{-1.5cm} \times f_{\sar}\left (\sar, M_\star, z\right ) \rm{dlog}\sar \rm{dlog} N_H 
\label{eq:effarea}
\end{eqnarray}


\noindent
where $\mathcal{I}\left (M_\star, \sar, z, N_H\right ) = \mathcal{I}(L_X,z,N_H)$ is the X-ray selection function given by the sensitivity map in the [2-10] keV band, $f\left (N_H\mid L_X, z\right )$ is the absorption distribution function from \citet{Ueda2014} and $f_{\sar}\left (\sar, M_\star, z\right )$ is the SARDF term in $\Psi_\sar(M_{\star},\sar,z)$. The latter term is required for the mass dependent incompleteness function $\mathcal{I}\left (M_\star, \sar, z, N_H\right )$ in addition to the ones needed for the computation of the luminosity dependent incompleteness function $\mathcal{I}(L_X,z)$.

While the V$_{max}$ method has the advantage of providing a non-parametric estimate of the AGN host galaxy mass function, it has the disadvantages that it requires a specific assumption for the SARDF term and, furthermore, it does not include the additional constraints from the AGN XLF, which, due to the limited luminosity  range probed by our sample, makes the results less robust in particular at the low mass end, where we only probe a limited range in \sar. On the contrary, the maximum likelihood provides a parametric estimate of the mass function, and determines the HGMF and SARDF simultaneously and self-consistently.
 Therefore we only use the V$_{max}$ method as a consistency check. For the function  $f_{ \sar\left (\sar, M_\star, z\right )}$ we assume the best fit $M_\star$-dependent SAR-term determined above (Eq.~\ref{eq:dpl}), normalized within $\log \sar>32\, [erg/s/M_{\odot}]$, which again defines our lower integration limit. 

The AGN Host Galaxy Mass Function is thus computed in three redshift bins as:

\begin{equation}
\Phi(M_\star) = \frac{1}{\Delta \rm logM_\star}\sum_{i=1}^{N_{obj}}  \frac{1}{V_{max}}
\end{equation}
and the binned values are shown in Fig.~\ref{fig:HGMF_ML_Vmax}, together with the Maximum likelihood result. 
The error bars are determined by bootstrapping of the sample with their $V_{\rm max}(M_\star)$ values.

As shown in the figure, we overall find a good agreement between the V$_{max}$ binned AGN HGMF and the AGN HGMF based on the maximum likelihood method. This confirms the  adopted parametric model in the maximum likelihood approach and verifies the robustness of our results.

\begin{figure*}
\begin{center}
\includegraphics[width=0.33\textwidth]{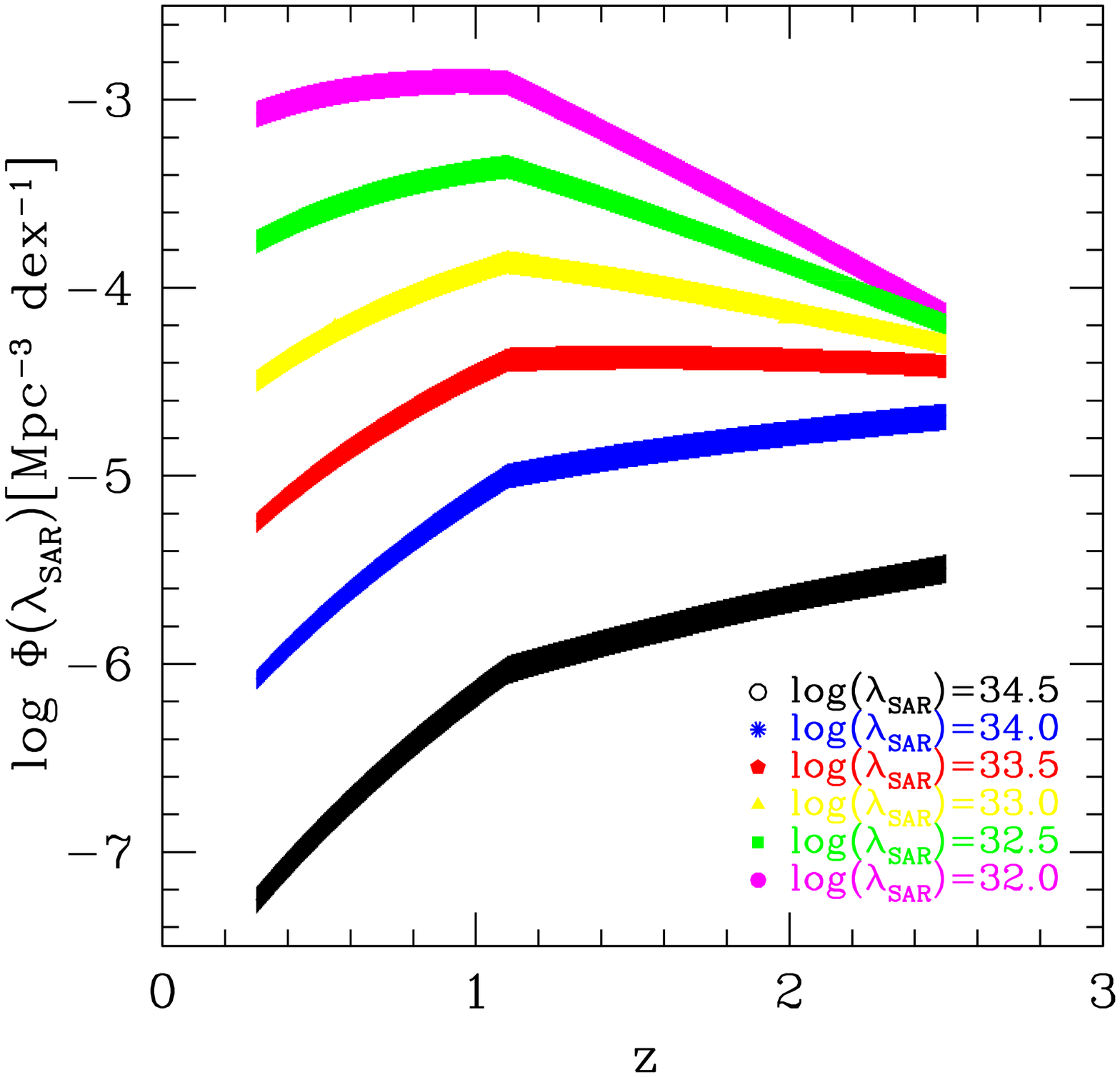}
\includegraphics[width=0.33\textwidth]{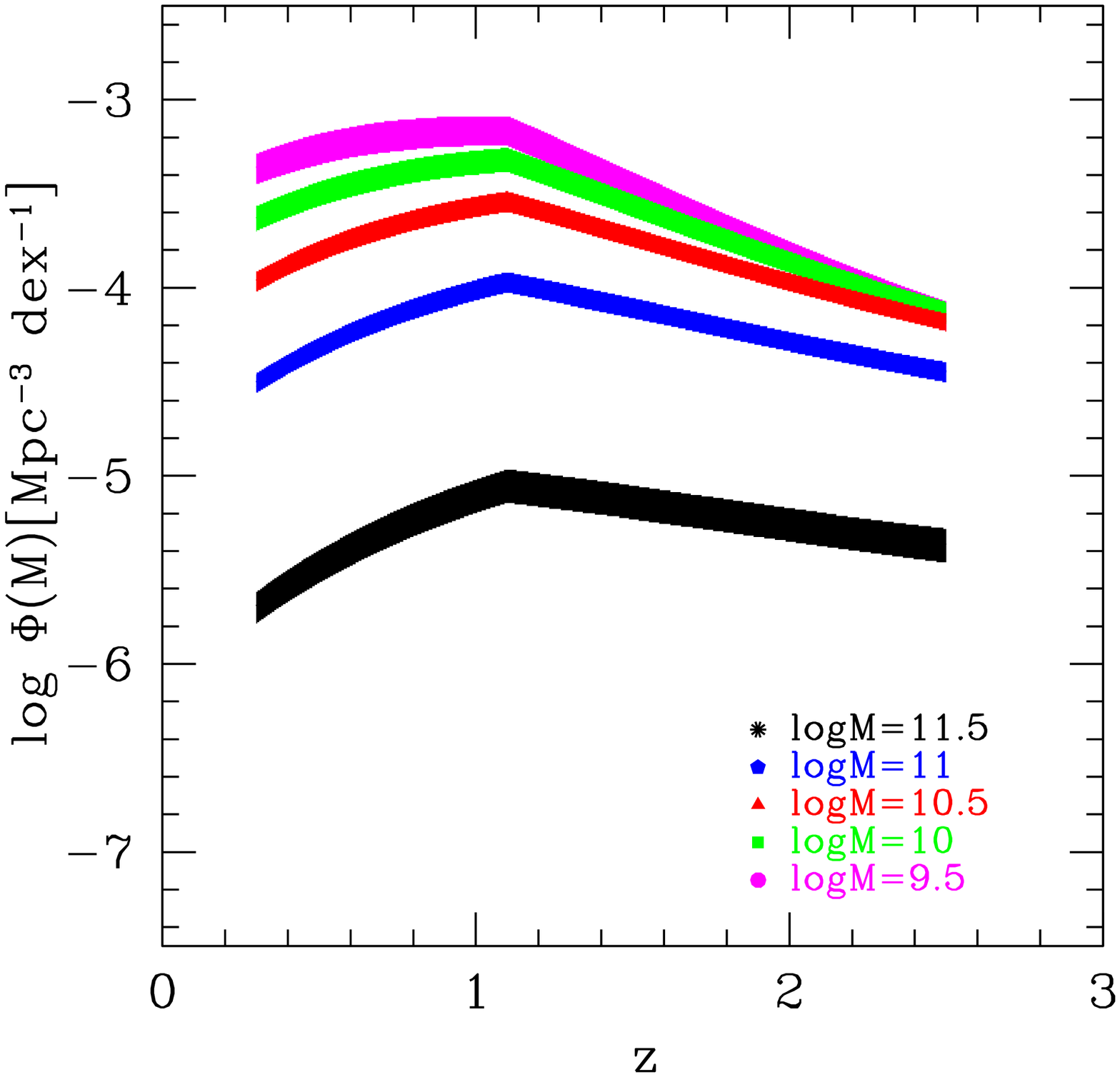}
\includegraphics[width=0.33\textwidth]{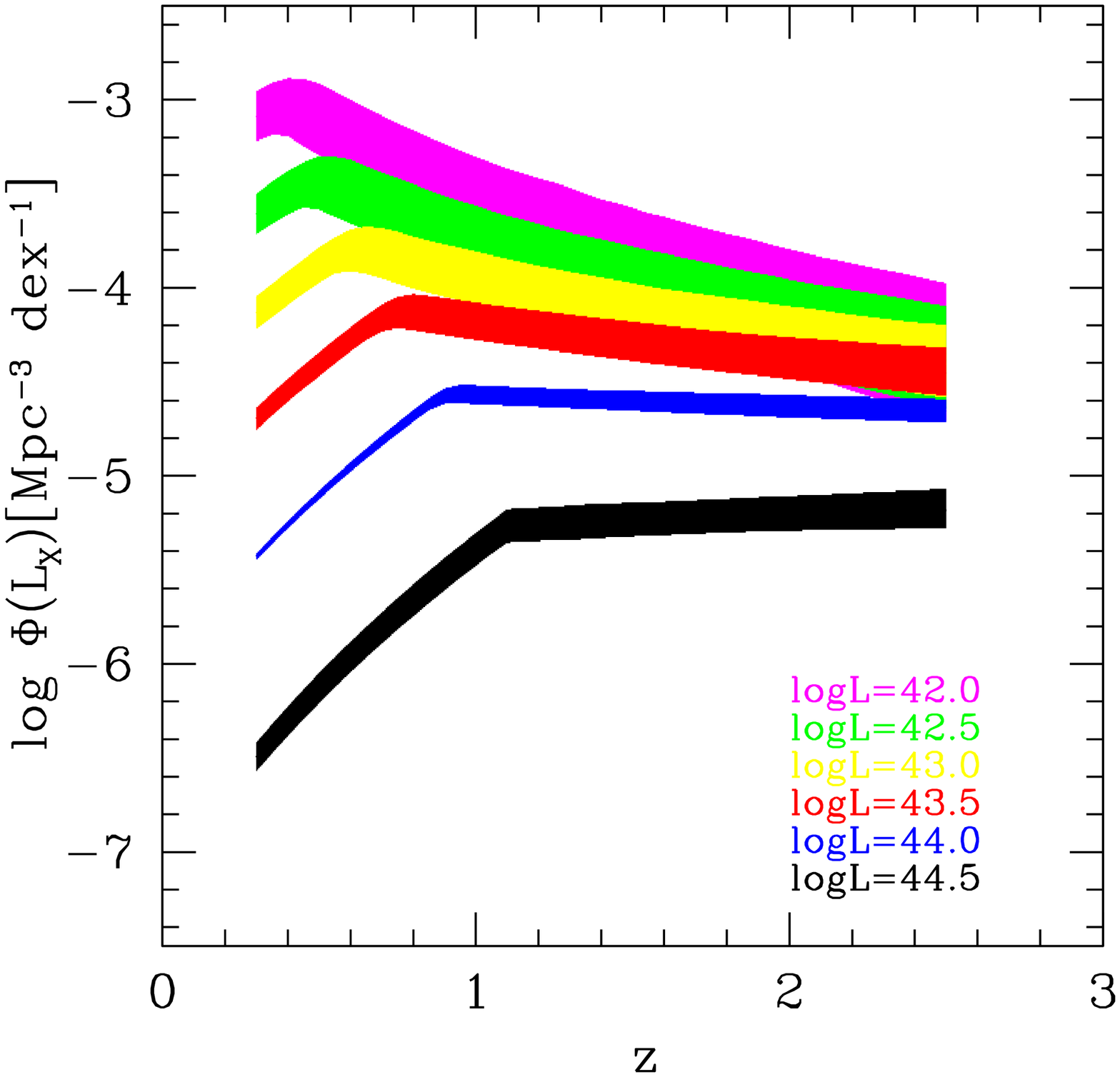}
\caption{\textit{Left Panel:} Redshift evolution of the SARDF space density for different \sar, i.e. magenta: log(\sar)=32$\,[erg/s/M_{\odot}]$ ($\sim$1\% Edd); green: log(\sar)=32.5$\,[erg/s/M_{\odot}]$ ($\sim$3\% Edd); yellow: log(\sar)=33$\,[erg/s/M_{\odot}]$ ($\sim$10\% Edd); red: log(\sar)=33.5$\,[erg/s/M_{\odot}]$ ($\sim$30\% Edd); blue: log(\sar)=34.0$\,[erg/s/M_{\odot}]$ ($\sim$ Edd); and black: log(\sar)=34.5$\,[erg/s/M_{\odot}]$ ($>$ Edd). \textit{Central Panel}: Redshift evolution of the HGMF space density for different M$_{\star}$. \textit{Right Panel}: Redshift evolution of the XLF space density for different $L_X$ by \citet{Miyaji2015}.
}
\label{fig:Edd_z}
\end{center}
\end{figure*}

\subsection{Results}
\label{sec:results}

In the upper panles of Fig. \ref{fig:Edd_ML}, we show the SAR-term f$_{\lambda_\sar}$ (Eq.~\ref{eq:dpl}), described by a double power-law with mass dependent, but redshift independent break $\lambda^*_{\rm SAR}$. The SARDF, shown in the lower panels of the same figure, is obtained by integrating the bivariate distribution function (including the above function) over M$_{\star}$. The SARDF can be described by a double power-law  whose low \sar\ characteristic slope flattens from -1.35 to -0.54 from the lowest to the highest redshift bin. The overall normalization $\phi^*_\lambda$ on the contrary increases for increasing redshift (see Tab.~\ref{tab:sarf}).
The increasing normalization with redshift was already noted in \citet{Aird2012} and \citet{Bongiorno2012}. In those works, the specific accretion rate distribution was parametrized with a single power law over the full redshift range, but already \citet{Bongiorno2012} noticed the presence of a break above log\sar$>34\,[erg/s/M_{\odot}]$. Furthermore, \citet{Aird2013} argued for a break in the specific accretion rate distribution to be consistent with the XLF.

While these previous studies do not report a change in the shape of the specific accretion rate distribution with redshift, we find a SARDF clearly flattening towards higher redshift.
It is important to note that, compared to the aforementioned works, there are some differences. First, here we determine the SARDF, i.e. the absolute space density as a function of $\sar$, while the previous studies present $p_\mathrm{AGN}(\sar | M_\star)$, i.e. the AGN fraction in the galaxy population. Furthermore, we account for obscuration by integration over the $N_H$ distribution, which generally steepens our low \sar\ slope.

The work by \citet{Aird2012} is refers to $0.2<z<1.0$, and thus it did not cover a sufficiently large redshift range to constrain this shape evolution. The sample used in \citet{Bongiorno2012} is instead similar and largely overlaps the one used in this study. A more accurate analysis of the sample used in \citet{Bongiorno2012} could indeed reveal the redshift dependence of the specific accretion rate, which was not included in the parametric model presented in \citet{Bongiorno2012}, due to the simpler single power-law parametrisation.
Finally, we model the bivariate distribution function of \sar\ and $M_\star$ and not only the SARDF and include additional information on the XLF. We discuss the effect of the latter in more detail in the Appendix.

\begin{figure*}
\begin{center}
\includegraphics[width=0.8\textwidth]{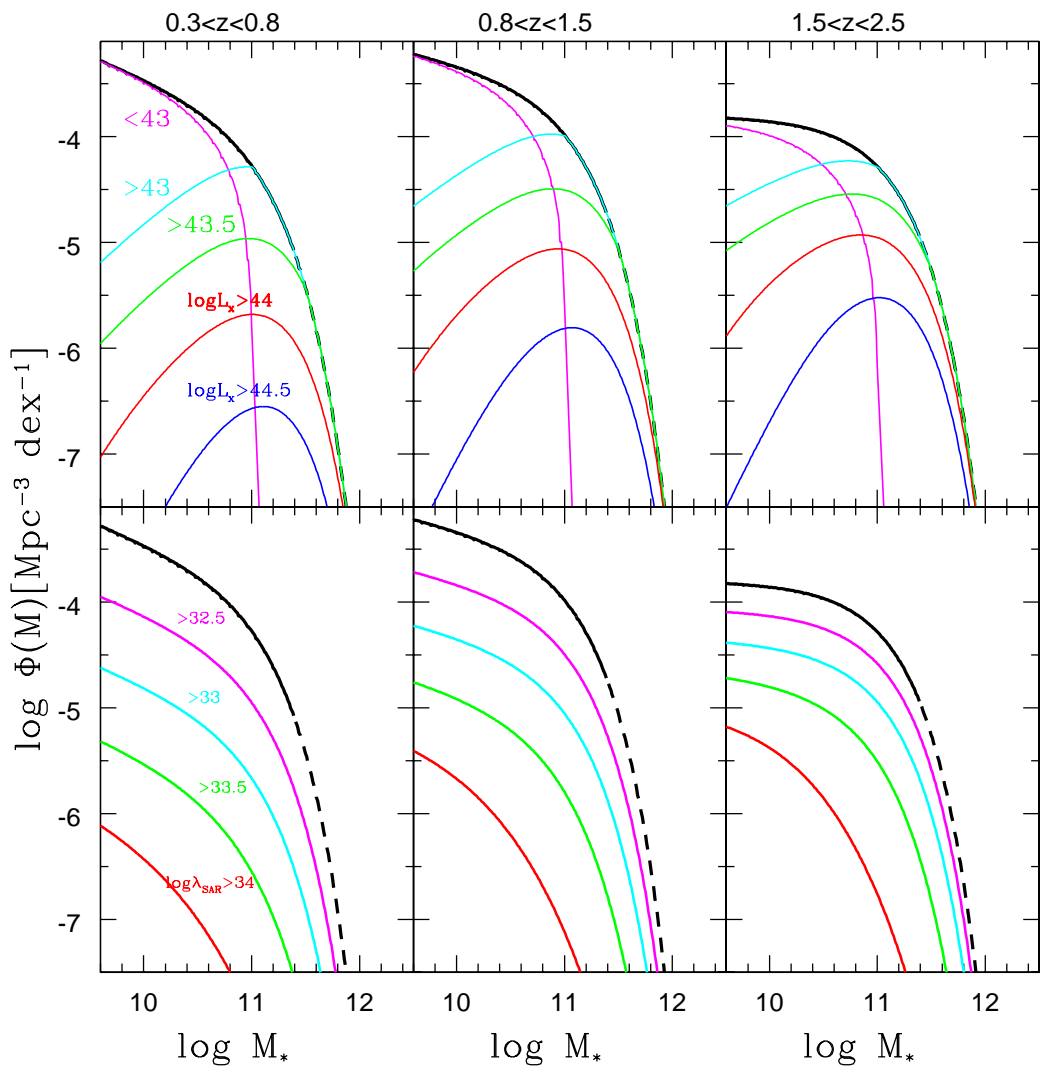}
\caption{Total AGN HGMF in three redshift bins derived through the  Maximum Likelihood (black line) and compared with the AGN HGMF for different AGN subsamples in (upper panels) luminosities, i.e log(L$_{X})<$43\,[erg/s] (magenta); log(L$_{X})>$43\,[erg/s] (cyan), log(L$_{X})>$43.5\,[erg/s] (green), log(L$_{X})>$44\,[erg/s] (red), and log(L$_{X})>$44.5\,[erg/s] (blue); and (lower panels) specific accretion rates \sar, i.e., log\sar$>32.5\,[erg/s/M_{\odot}]$ (magenta); log\sar$>33\,[erg/s/M_{\odot}]$ (cyan), log\sar$>33.5\,[erg/s/M_{\odot}]$  (green), and log\sar$>34\,[erg/s/M_{\odot}]$ (red).}
\label{fig:HGMF_L_Edd_cuts}
\end{center}
\end{figure*}

The best fit HGMF (black line in Fig. \ref{fig:HGMF_ML_Vmax}) is well described by a Schechter function with constant $\rm M^*_{\star}$ and a low mass slope $\alpha$ flattening with redshift (i.e. $\alpha=-0.41$
in the first redshift bin, $-0.24$ in the second and $-0.03$ in the third one; see Eq.~\ref{tab:mf}). 
We compare the AGN HGMF with the total galaxy stellar mass function (red curve and shaded magenta region) and the star forming galaxy mass function (blue curve and shaded cyan region) by \citet{Ilbert2013}.
We note that, at $\rm log(M_{\star}/M_{\odot})>$11.5, the HGMF but also the total and SF galaxy mass functions are not well constrained by the data (see Fig.~\ref{fig:sample}) due to the limited volume sampled in both cases. This region is indicated by the dashed lines in Fig.~\ref{fig:HGMF_ML_Vmax}. Furthermore, in the highest $z$-bin the galaxy mass function of \citet{Ilbert2013} shows an upturn at low masses, captured in their double Schechter function model, which is not captured in our more restricted single Schechter function model for the HGMF. Our data do not allow to constrain such an upturn for our AGN sample, which would require a larger sample, and probably a deeper flux limit for the galaxies including lower luminosity AGN. 

The ratio of AGN HGMF over total galaxy mass function is shown by the red line and the shaded magenta area in the lower panels of Fig. \ref{fig:HGMF_ML_Vmax}. Such ratio indicates the active fraction or duty cycle of AGN activity in the galaxy population, if we consider AGN with $\log \sar>32$ ($\sim 1$\% Eddington) which corresponds to the definition of an AGN assumed in this paper. 

We find a redshift evolution in the mass dependence of the active fraction. At $M_\star=10^{10}\,M_\odot$, the active fraction is  approximately constant at $\sim10\%$, while at $M_\star=10^{11.5}\,M_\odot$ it increases over our three redshift bins from $\sim 3\%$ to $\sim 8\%$ to $\sim 20\%$. This trend is in qualitative agreement with the results for the SMBH mass dependence of the active fraction of the black hole mass function, presented in \citet{Schulze2015}. This could be related to the redshift evolution of the gas reservoir available to fuel the AGN, since in high redshift galaxies a greater amount of gas can be responsible for triggering AGN activity \citep{Tacconi2010}.

The ratio of AGN HGMF to the star forming mass function (shown by the blue line in the lower panels of Fig. \ref{fig:HGMF_ML_Vmax}) traces the average relation between star forming and AGN activity as a function of stellar mass. It extends the well known average agreement between star formation rate density and black hole accretion density \citep[e.g.][]{Marconi2004} to its stellar mass dependence. Overall we find a weaker redshift evolution in the shape of this ratio than for the active fraction, where the ratio stays almost constant over $10^{10}<M_\star<10^{11}\,M_\odot$, the mass range tracing the bulk of the population, in all three redshift bins. At the high mass end for $z>0.8$ the AGN/SF galaxy ratio and for $z>1.5$ also the active fraction appear to increase with stellar mass. Future studies will be required to confirm or disprove the reality of this trend.

The redshift evolution of the SARDF and HGMF allows a more detailed look at the AGN downsizing behaviour, i.e. the luminosity-dependent evolution, seen in the XLF out to $z\sim2.5$. They probe the more physically meaningful quantities stellar mass and specific accretion rate distribution, and by inference relate to black hole mass and Eddington ratio. In Fig. \ref{fig:Edd_z} we show the global trend of the redshift evolution of the space density in bins of \sar\ (left panel), $M_\star$ (central panel) and $L_X$ (right panel). The $L_X$ dependence, based on the XLF from \citet{Miyaji2015} shows the well known AGN downsizing behaviour \citep[e.g.][]{Ueda2003,Hasinger2005,LaFranca2005,Bongiorno2007,Silverman2008}. For the \sar\ dependence, we see that higher \sar\ objects ($\log\sar>33.5\,[erg/s/M_{\odot}]$) have a peak in their space density at an earlier cosmic epoch compared to the lower \sar\ objects ($\log\sar<33.5\,[erg/s/M_{\odot}]$), i.e. also showing a clear downsizing trend. The $M_\star$ dependence, based on the HGMF, also indicates a downsizing trend, with AGN in lower stellar mass galaxies showing a steeper decline in their space density towards high redshift than higher stellar mass galaxies, but less pronounced than what is seen in the SARDF. This suggests that the downsizing in the AGN luminosity function is due to the combination of a (weak) mass-dependent evolution of the HGMF and the stronger evolution of the SARDF.

In Fig.~\ref{fig:HGMF_L_Edd_cuts} upper panels, we show the AGN HGMF for different luminosity  sub-classes, i.e.  $\rm \log(L_{X})<$43\,[erg/s] (magenta), $\rm \log(L_{X})>$43\,[erg/s] (cyan), $\rm \log(L_{X})>$43.5\,[erg/s] (green), $\rm \log(L_{X})>$44\,[erg/s] (red), and $\rm \log(L_{X})>$44.5\,[erg/s] (blue).
As expected the high mass end is dominated by luminous AGN ($\rm logL_X>43$\,[erg/s]), while the low mass bins are mainly populated by low luminosity objects ($\rm logL_X<43$\,[erg/s]) whose contribution above $\rm log(M_{\star}/M_{\odot})\sim$11 is negligible. Our definition threshold of $\log L_X/M_\star>32$ directly excludes any AGN with $\log L_X<43$\,[erg/s] above $M_\star>10^{11}\,M_\odot$. This also implies that when applying an AGN definition by a luminosity threshold, as usually done, you will tend to find an active fraction increasing with mass, consistent with previous work \citep[e.g.][]{Bundy2008,Xue2010,Aird2012,Silverman2009}.

In the lower panels, we instead show the total AGN HGMF in \sar\ bins i.e.,  log\sar$>32.5\, [erg/s/M_{\odot}]$ (magenta), log\sar$>33\, [erg/s/M_{\odot}]$ (cyan), log\sar$>33.5\, [erg/s/M_{\odot}]$  (green), and log\sar$>34\, [erg/s/M_{\odot}]$ (red). Overall, the mass distributions of AGN of different specific accretion rate have a similar shape, only mildly affected by the $M_\star$ dependence in our SARDF model.

 \begin{figure*}
 \begin{center}
\includegraphics[width=0.8\textwidth]{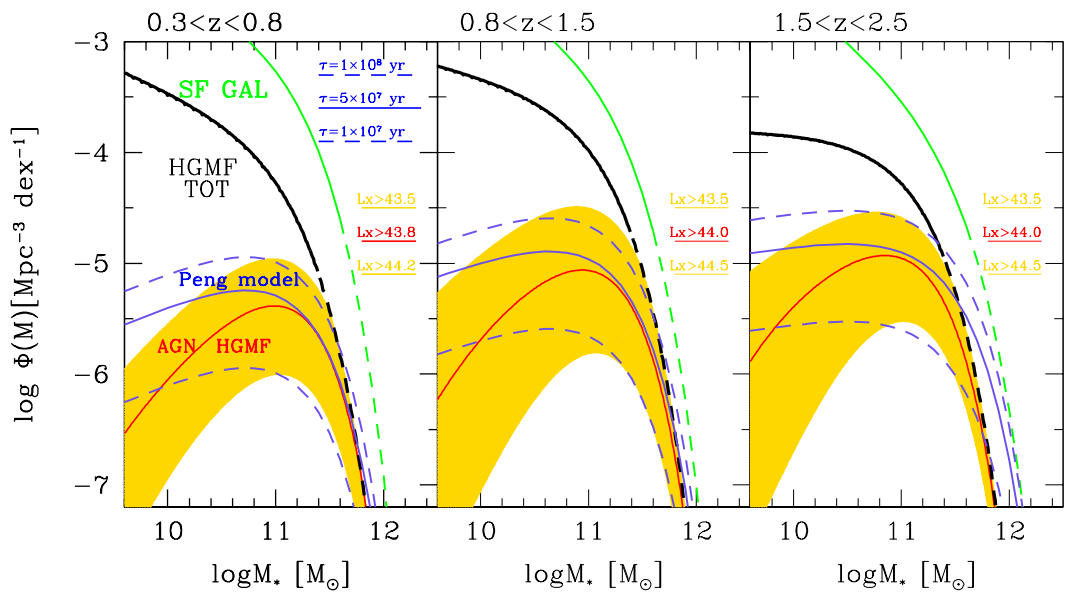}
\caption{Predicted MF of transient galaxies as derived starting from a single Schechter function based on the data from \citet[][green line]{Ilbert2013}, using the Peng recipe \citep{Peng2010} for $\tau=5\times10^7$yr (blue solid line), $\tau=1\times10^7$yr and $\tau=1\times10^8$yr (blue dashed lines). We compare them with the total AGN HGMF (black line) and with the HGMF computed with different luminosity cuts. In particular in the first redshift bin, cuts of $\rm logL_X>$43.8\,[erg/s] (red solid line) and $\rm logL_X>$44.2 and $\rm logL_X>$43.5\,[erg/s] for the lower and upper boundary, respectively have been applied; while in the second and third redshift bins, these cuts are $\rm logL_X>$44 (brown solid line) and $\rm logL_X>$44.5\,[erg/s] and $\rm logL_X>$43.5\,[erg/s] for the lower and upper boundaries.}
\label{fig:mf_quench}
\end{center}
\end{figure*}

\section{The mass function of galaxies in the process of being mass-quenched}
 \label{sec:quenching}
 
According to the model described in \citet{Peng2010}, the quenching process, i.e. the process which leads to the transition from star-forming to passive galaxies, independent of its physical origin, can be described by two different modes: mass and environment quenching, whose differential effects on the fraction of passive/red galaxies are separable.

In \citet{Peng2010} paper, it is speculated that the environment quenching occur in satellite galaxies, while the mass quenching could reflect a feedback mechanism related to star-formation or AGN. 
In a subsequent paper, \citet{Peng2012} confirm the expectation on the environment quenching as due to satellite galaxies, studying the mass function of central and satellite galaxies. 
Here we want to test whether the mass quenching process can be linked to AGN feedback. 

The strength of the \citet{Peng2010} approach is that this phenomenological model is based on simple observational inputs, which allow one to successfully reproduce many of the features of the galaxy population. Moreover, the model is able to give a clear prediction for the mass function of the galaxies in the process of being mass-quenched and the inter-relationships between the Schechter parameters for star-forming and passive galaxies. 

The mass function of the transient population can be described by a single Schechter function with parameters \citep[see eq. (28) of][]{Peng2010}:
\begin{equation}
\label{eq:peng}
\begin{split}
 &\rm M^*_{\star,trans} = M^*_{\star,blue} \\
  &\rm \alpha_{s,trans} = \alpha_{s,blue} + (1+\beta) \\
  &\rm \Phi^{*}_{trans} = \Phi^*_{blue} \, sSFR(M_\star,z)|_{M^*} \, \tau_{trans} \\
\end{split}
\end{equation}

\noindent
where $M^*_{\star,blue}$, $\alpha_{s,blue}$ and $\Phi^*_{blue}$ are the parameters of the Schechter function which describes the star-forming galaxy mass function and $\beta$ is the exponent in the power law relation that links the specific star formation rate (sSFR) and the stellar mass (see Eq.~\ref{eq:sSFR}). Here we use the data for star-forming galaxies from \citet{Ilbert2013}, and force the fit with  a single Schechter function. This parametric choice is required to use the model fits provided by \citet{Peng2010} with a single Schechter function as starting MF. This introduces some uncertainties especially with respect to slope of the high-mass end, which is the most difficult part of the stellar MF to be constrained, as we will point out later in this section. The value $\tau_{trans}$ is the period of time the ``transient'' signature is visible, and is not constrained by the \citet{Peng2010} model. Here, we assume that the transient phase corresponds to the active feedback/blow-out phase i.e. the gas depletion time-scale associated with the outflow. Current observations suggest this time-scale to be of the order of $\rm 1-10\times10^7 yr$ \citep{Maiolino2007,Feruglio2010,Cicone2014}. 
Finally, $\rm sSFR(M_\star,z)$ is the evolving specific star formation rate. Here we consider the recent measurement of the $\rm sSFR$ from \citet[][eq. (2)]{Lilly2013}:

\begin{equation}
\begin{aligned} 
\rm sSFR(M,z)=& \rm 0.07 \left(\frac{M_\star}{10^{10.5} M_{\odot}} \right)^{\beta}   \left (1+z\right)^{3} Gyr^{-1} &\rm for \, z<2 \\
\rm sSFR(M,z) = & \rm 0.30 \left(\frac{M_\star}{10^{10.5} M_{\odot}} \right)^{\beta}   \left (1+z\right)^{5/3} Gyr^{-1} &\rm for \, z>2\\
\label{eq:sSFR}
\end{aligned}
\end{equation}

\noindent
with $\beta \sim -0.1$.

Starting from the star-forming galaxy mass function (green line in Fig.~\ref{fig:mf_quench}), we then derive, using the above equations,  the predicted mass function  of the transient, i.e. ``in the process of being mass-quenched'', population. Given the uncertainties on the value of $\tau_{trans}$ we show in Fig.~\ref{fig:mf_quench} the predictions for a range of $\rm \tau_{trans}=1-10 \times 10^7$; the blue solid line is for $\rm \tau=5\times 10^{7}$ while the blue dashed lines correspond to $\rm 1\times 10^7 yr$ and $\rm 1 \times 10^8 yr$ (lower and upper boundary, respectively). 

To test whether AGN can be responsible for the mass-quenching of galaxies, we chose to restrict our analysis to the most luminous objects.
Theory indeed predicts that the  capability of AGN outflows of perturbing the ISM depends on AGN luminosity as $L^{1/2}_{\rm Bol}$ \citep{Menci2008} and  that galaxy-scale outflows are energy-driven, i.e., their mechanical energy is proportional to the AGN  luminosity \citep{Zubovas2012}.
This scenario is supported by observations that find that the momentum rate of kpc-scale outflows \citep{Sturm2011,Cicone2014,Feruglio2015} is $\geq10-20\,L_{Bol}/c$, i.e. the more luminous the AGN is, the more powerful outflows are produced. This means that the AGN-driven feedback mechanism should become increasingly more efficient in halting the star-formation in  the host galaxy for higher AGN luminosities.

In Fig.~\ref{fig:mf_quench} we compare the prediction for the mass function of mass quenching transient objects with the HGMF of the total population, i.e. $\log \sar>32$, and of different sub-samples. We test the agreement using sub-samples applying in addition different cuts on either $L_X$ or \sar, as shown in Fig.~\ref{fig:HGMF_L_Edd_cuts}. We do not consider more complicated cuts or for example a luminosity dependent transition time-scale, which could improve the agreement between the two mass functions, in order to keep the comparison as simple as possible.
We find that the class of objects that best reproduces, in terms of both shape and normalization, the expected mass function are: $\rm logL_X>43.8^{+0.4}_{-0.3}$\,[erg/s] (red solid line and yellow shaded area) at $\rm 0.3<z<0.8$, and $\rm logL_X>44\pm0.5$\,[erg/s] at $\rm 0.8<z<2.5$. Reducing the threshold in $\rm L_X$ leads to a space density in the HGMF higher than the expected for the "transient" objects at the low mass end.
On the contrary, specific accretion rate based sub-samples do not seem to reproduce the expected mass function particularly well. This is because, within the \citet{Peng2010} model, for a constant $\tau_{trans}$,  the fractional density of the "transition" population strongly decreases at low masses: only very few low-mass galaxies experience quenching at any redshift. On the other hand, the poluation of AGN above any given \sar\ threshold increases towards low stellar masses (see the bottom panel of Fig.~\ref{fig:HGMF_L_Edd_cuts}): rapidly growing high Eddington ratio objects can be found in galaxies of any mass, at all redshifts. Thus, any model that invokes a fixed threshold in \sar\ to explain the quenching population \citep[see e.g.][]{Zubovas2012} would predict a too high fraction of low-mass galaxies in the transition phase, in strong contrast with the \citet{Peng2010} finding.

We note that the disagreement between the Peng model prediction and the AGN HGMF present at the high mass end of the third redshift bin is due to the fact that, to apply the recipe from \citet{Peng2010}, we used a single Schechter function to fit the star-forming galaxy mass function which, as pointed out by \citet{Ilbert2013}, is not a good fit of the data, especially at the high mass end. The fit with a double Schechter function (as performed in \citet{Ilbert2013} and shown in Fig.~\ref{fig:HGMF_ML_Vmax}) would indeed be steeper thus reducing the number of predicted high mass objects and the discrepancy with the AGN HGMF.

Overall we find the space density of luminous AGN ($\rm logL_{\rm X}>43.5-44.5$\,[erg/s]) at stellar masses $M_\star>10^{10.7} M_{\odot}$ to be consistent with the space density of galaxies in the star formation quenching phase. 
This non trivial result is consistent with the notion that feedback from luminous AGN can be associated to the mass-quenching of galaxies. 
At lower masses the difference in space density between the luminous AGN mass function and the quenching mass function leaves room for a contribution via another mechanism. Lower luminosity AGN might contribute here, if their AGN feedback mechanism would operate on a different transition time-scale $\rm \tau_{trans}$.
Furthermore, \citet{Peng2015} recently suggested that ``strangulation'' (a mechanism for which the supply of cold gas is halted) is the primary mechanism responsible for quenching star-formation in local galaxies with a stellar mass less than 10$^{11}$M$_{\odot}$. 
Our results are complementary to this work, proposing AGN-driven outflows as a plausible mechanism for halting star formation at higher redshift and for more massive galaxies, although a causal connection is not substantiated.

\section{Summary and Conclusions}

In this paper, we have studied the host galaxy stellar mass function of a sample of $\sim$1000 AGN detected in the XMM-COSMOS field in the 2-10keV band at 0.3$<z<$2.5. 
We derived the SARDF and the HGMF simultaneously as a bivariate distribution function of stellar mass and specific accretion rate \sar $\equiv  L_X/M_\star$, using the maximum likelihood method developed by \citet{Schulze2010} and extended by \citet{Schulze2015}. 

\noindent
Our results can be summarized as follows:
\begin{itemize}
\item[(i)]The SARDF is best described by a double power-law with a mass dependent but redshift independent break $\lambda^*_{\rm SAR}$ and  a low \sar\ characteristic slope which flattens from $-1.35$ to $-0.54$ with increasing redshift. The overall normalization $\phi^*_\lambda$ on the contrary increases for increasing redshift. 

\item[(ii)] The AGN HGMF is described by a Schechter function with constant $\rm M^*_{\star}$ and a low mass slope $\alpha$ flattening with redshift from $\alpha$=-0.41 at $z=0.55$ to $\alpha$=-0.03 at $z=2.0$. 
We derived the active fraction of AGN activity by comparison with the stellar mass function by \citet{Ilbert2013} and we find a redshift evolution in its mass dependence at the high mass end, where the the fraction of AGN in massive galaxies  increases from $\sim$3\% at $z\sim0.55$ to $\sim$20\% at $z\sim2$.

\item[(iii)] The redshift evolution of the SARDF and AGN HGMF allows us to gain a deeper understanding into the physical drivers of the AGN downsizing behaviour, seen in the XLF out to $z\sim2.5$. We find that that the downsizing in the AGN luminosity function is due to the combination of a (weak) mass-dependent evolution of the HGMF and the stronger evolution of the SARDF. In particular, we see that higher \sar\ objects have a peak in their space density at earlier epoch compared to the lower \sar\ AGN.

\item[(iv)] We compare the mass function of the population in the process of being ``mass-quenched'', predicted by the phenomenological model by \citet{Peng2010}, with the HGMF computed for different sub-samples obtained with different luminosity and \sar-cuts. We find at the high masses (i.e.  $M_\star>10^{10.7} M_{\odot}$) that the population that agrees with the model prediction is that of luminous AGN having $\rm logL_{\rm X}>43.5-44.5$ \,[erg/s] (i.e. $\rm logL_{bol}\gtrsim46$\,[erg/s]). Both their number density and stellar mass distribution are consistent with those of the ``transition'' galaxy population, a crucial, and non trivial, result of our analysis. While this agreement does not establish a causal connection between star formation quenching and AGN activity, it suggests AGN feedback by powerful outflows from luminous AGN as a plausible mechanism for the mass-quenching of star forming galaxies. This scenario would be in agreement and complementary with the recent findings by \citet{Peng2015} who suggested ``strangulation'' as the primary quenching mechanism at lower masses (i.e. $M_\star<10^{11} M_{\odot}$).

\end{itemize}

\begin{acknowledgements}
We thank the referee for the careful reading and precious suggestions which helped to improve the manuscript. This work is based on the COSMOS program.
The HST COSMOS Treasury program
was supported through NASA grant HST-GO-09822. This work is mainly based on observations obtained with XMM-Newton,
an ESA Science Mission with instruments and contributions directly funded by ESA Member States and the USA (NASA),
and with the European Southern Observatory under Large Program 175.A-0839, Chile. 
A.B. and E.P. acknowledge financial support from INAF under the contract
PRIN-INAF-2012
A.S. acknowledges support by JSPS KAKENHI Grant Number 26800098. MCMC corner plots make use of \texttt{triangle.py} \citep{Foreman2013}.

\end{acknowledgements}

\appendix

\section{Model comparison} \label{app:a}

\begin{figure*}
\centering
\includegraphics[width=0.9\textwidth]{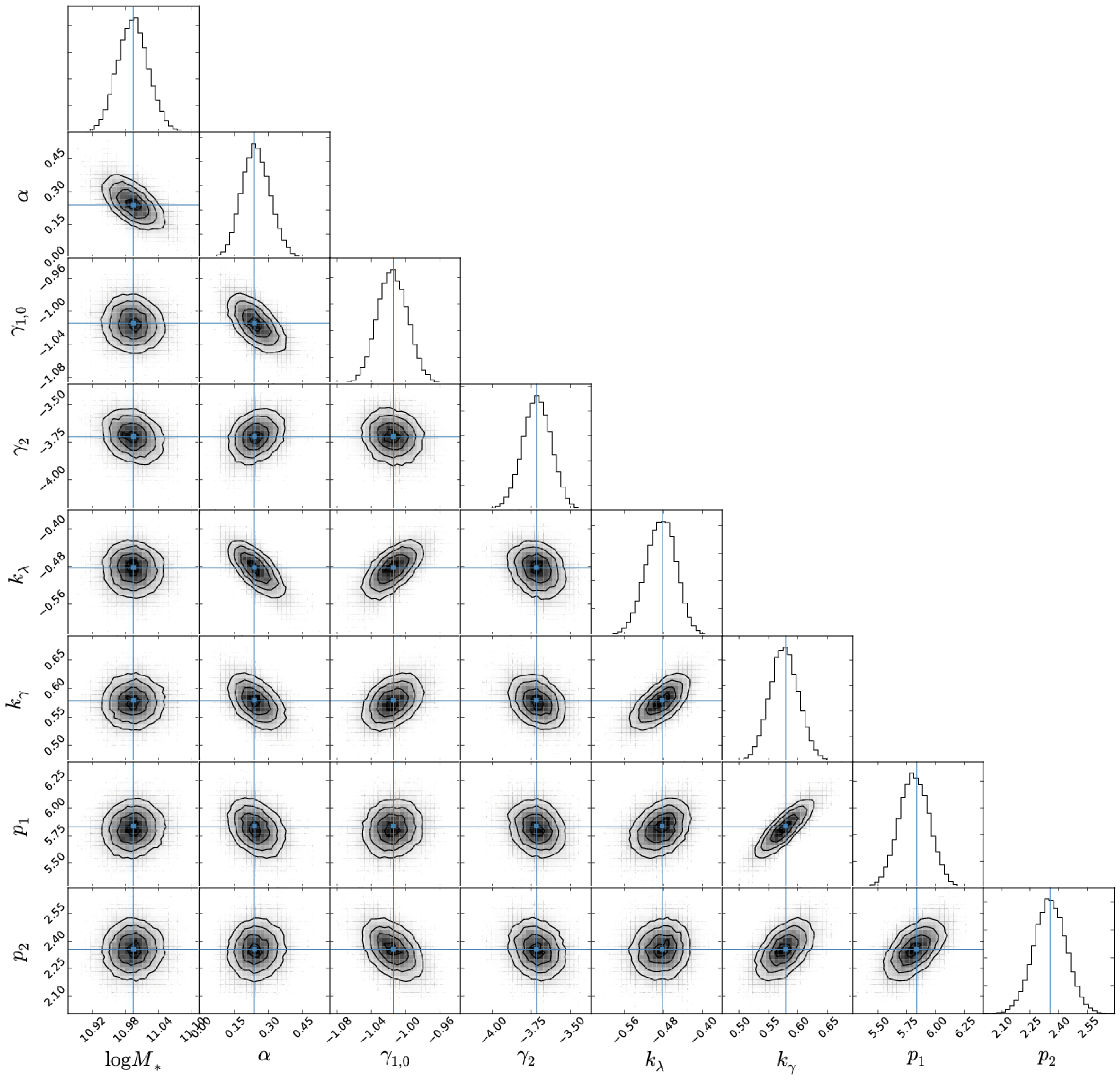}
\caption{Posterior probability distribution for the free parameters of our best fit model. The histograms on the top show the marginalized probability distribution function for each free parameter, while the contour plots show the marginalized 2-dimensional probability distribution function for each pair of parameters, to illustrate their covariances.}
\label{fig:triangle}
\end{figure*}

\begin{figure*}
\centering
\includegraphics[width=0.9\textwidth]{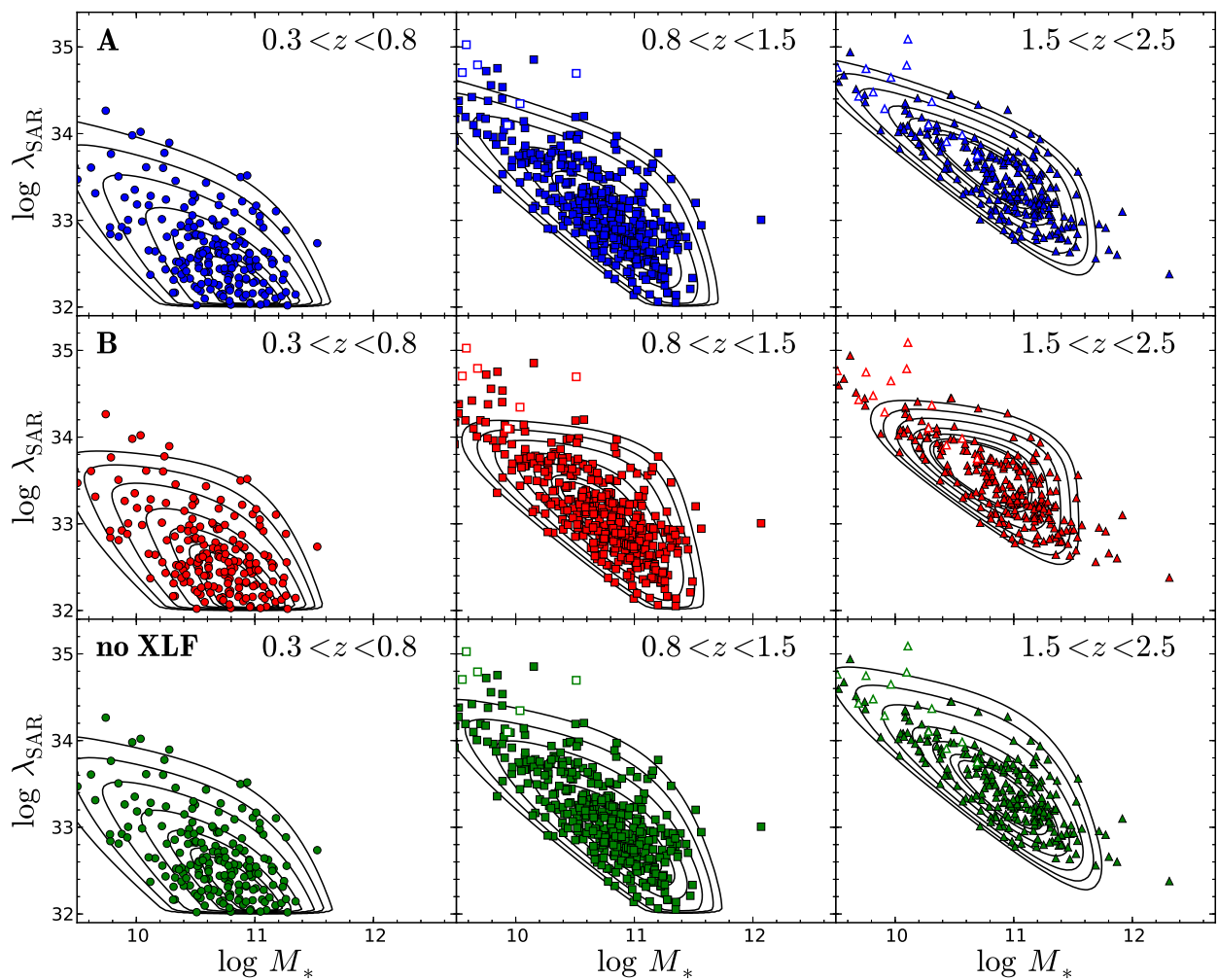}
\caption{Comparison of the bivariate distribution  in the M$_\star-\sar$ plane between the observations (symbols) and the prediction from the best fit model (black contours) in the three redshift bins as labeled. The upper panels show the comparison for our default model (A), the middle panels show the prediction from the best fit  $M_\star$-independent SARDF model (B), and the lower panels show the best fit solution if no XLF data is included in the likelihood function ("no XLF").
Open symbols indicate upper limits in $M_\star$.}
\label{fig:msarplane}
\end{figure*}

\begin{figure*}
\centering
\includegraphics[width=0.9\textwidth]{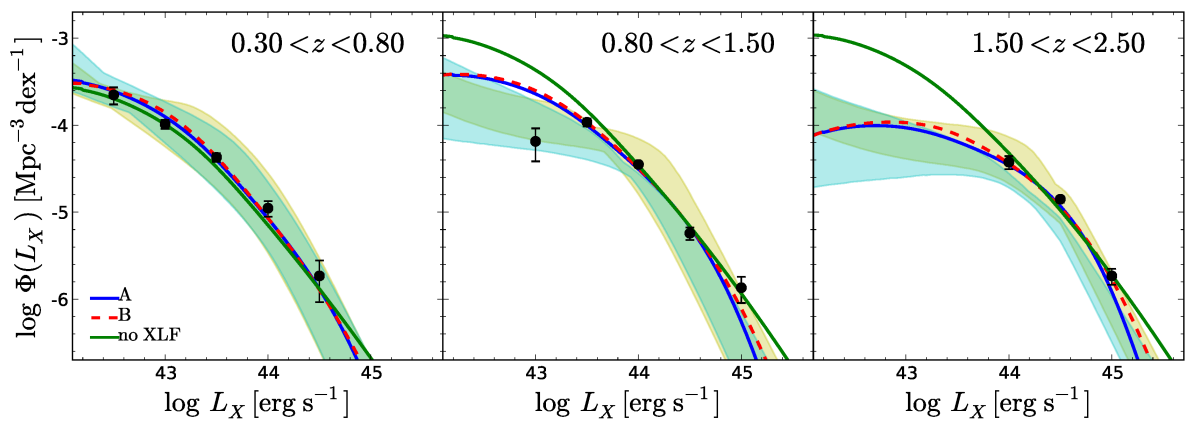}
\caption{Comparison of the X-ray luminosity function (XLF) predicted by the bivariate distribution $\Psi(M_\star,\sar,z)$ with direct observations. The blue solid line shows the XLF derived from our default model (A), the red dashed line is for the best fit  $M_\star$-independent SARDF model (B) and the green solid line is for the best fit without including the XLF into the likelihood function (no XLF). The black circles show the binned XLF for our XMM-COSMOS sample using the $V_\mathrm{max}$ method, indicating the luminosity range covered by XMM-COSMOS. The shaded cyan and yellow areas show the XLF by \citet{Miyaji2015} and \citet{Ueda2014} respectively, where the shaded area includes the variation of the XLF over the redshift range of the bin and the uncertainty of the XLF determination.}
\label{fig:xlf}
\end{figure*}

\begin{table*}
\centering
\begin{tabular}{c c |c c c  | c c c c c c | c c c}
\hline \hline 

\multicolumn{2}{ c |}{} &\multicolumn{3}{ c |}{}&\multicolumn{6}{c |}{}&\multicolumn{3}{c }{}\\
Model & $\log (\Psi^\ast)$  & $\log M_\star^*[M_{\odot}]$ & $\alpha_0$ & $k_\alpha$ & $\log\lambda^*_{\rm SAR}$ & $k_\lambda$ & $\log M_{\star,0}$ &$\gamma_{1,0}$ & $k_\gamma$  & $\gamma_2$ & $p_1$ & $p_2$ & $z_0$ \\ 
\hline 
\multicolumn{2}{ c |}{}&\multicolumn{3}{c |}{}&\multicolumn{6}{c |}{}&\multicolumn{3}{c }{}\\
no XLF & -6.69 & 11.02 & -0.503 & $0.0^*$ & $33.8^*$ & 0.05 & $11.0^*$ & -1.11 & 0.17 & -1.97 & 5.49 & 1.45 &  $1.1^*$  \\
B & -6.67 & 10.88 & -0.192 & $0.0^*$ & $33.8^*$ & $0^*$ & $11.0^*$ & -0.85 & 0.82 & -3.35 & 6.32 & 2.39 & $1.1^*$  \\ 
C & -6.90 & 11.00 & 0.225 & -0.017 & $33.8^*$ & -0.48 & $11.0^*$ & -1.01 & 0.60 & -3.80 & 5.90 & 2.37 & $1.1^*$  \\ 
\hline
\hline
\end{tabular}
\vspace{0.2cm}
\caption{Best fit model parameters for the bivariate distribution function of stellar mass and SAR for other models, which have been tested but where not considered further.
The parameters denoted with an $^*$ are fixed during the fit.}
\label{tab:modelB}
\end{table*}

 \begin{table*}
\centering
\begin{tabular}{lccc}
\hline \hline \noalign{\smallskip}
Model & fixed parameters & $\Delta$AIC$_c$ & relative likelihood \\
\hline \noalign{\smallskip}
default (A) & $k_\alpha=0$ / $k_\lambda$, $k_\gamma$ free & 0 & 1.0 \\ \noalign{\smallskip}
(A) + $z$-evolution in $M_\star^*$ & $k_\alpha=0$ / $k_\lambda$, $k_\gamma$, $k_{M_\star^*}$ free & 10 & $5.5\times10^{-3}$    \\ \noalign{\smallskip}
$z$-evolution in $\alpha$ (C) & $k_\alpha$, $k_\lambda$, $k_\gamma$ free & 14 & $9.5\times10^{-4}$ \\ \noalign{\smallskip}
(C) + $z$-evolution in $M_\star^*$ & $k_\alpha$, $k_\lambda$, $k_\gamma$, $k_{M_\star^*}$ free & 28 & $1.0\times10^{-6}$ \\ \noalign{\smallskip}
no $M_\star$ dependence in $f_\sar$ (B) & $k_\alpha=0$, $k_\lambda=0$ / $k_\gamma$ free & 311 & $2.7\times10^{-68}$    \\ \noalign{\smallskip}
(B) + $z$-evolution in $\alpha$ & $k_\lambda=0$ / $k_\alpha$, $k_\gamma$ free & 311 & $2.5\times10^{-68}$    \\ \noalign{\smallskip}
no $z$-evolution in $f_\sar$ & $k_\gamma=0$ / $k_\alpha$, $k_\lambda$ free & 531 & $6.4\times10^{-116}$ \\ \noalign{\smallskip}
no $z$-evolution in $f_\sar$ and $f_\star$  & $k_\alpha=0$, $k_\gamma=0$ / $k_\lambda$ free & 731 & $1.7\times10^{-159}$ \\
\hline
\hline
\end{tabular}
\vspace{0.2cm}
\caption{Comparison between different parametric models and our default model via their difference in AIC$_c$ and their corresponding relative likelihood.}
\label{tab:aic}
\end{table*}

As mentioned in Sec.~\ref{sec:ML}, we derived uncertainties via MCMC computation of the posterior distribution function (PDF). In Fig.~\ref{fig:triangle} we show the 1D marginalized PDF for the free parameters of our model and the 2D marginalized PDF for parameter pairs. The latter shows the covariance between these pairs. We find covariance between several parameters, e.g. between the break $\log M_\star^*$ and the 
slope $\alpha$ of the $M_\star$-term $f_\star$ or between several of the redshift evolution parameters ($p_1$, $p_2$, $k_\gamma$).

Besides the default parametric model we presented above, we also explored additional parametric models for the HGMF and SARDF. We here discuss the results of this exercise and provide a justification for our chosen parameterization. To compare the relative quality of our respective parametric model given our data set,  we use the Akaike information criterion \citep[AIC$_c$;][]{Akaike1974}. It is given by AIC$_c=S+2K+2K (N / ( N - K - 1))$, where $S$ is the likelihood as defined above, $K$ is the number of parameters in the model and $N$ is the size of the sample.
While the AIC$_c$ penalizes against overfitting, it is known to be less penalizing than e.g. the Bayesian information criterion (BIC). Nevertheless, in general the difference in AIC between the models tested below is significant enough to draw firm conclusions.

Our default model (hereafter model A) allows for a mass dependence in the SARDF, while previous studies assumed a \sar\ distribution, independent of mass \citep[e.g.][]{Aird2012, Bongiorno2012}. In fact, we first started our analysis with a model without such a mass dependent term, i.e. $k_\lambda=0$ (hereafter model B). Our best fit model B provides an almost equivalent fit to the XLF as our default model A (see red dashed line in Fig.~\ref{fig:xlf}) and fits well over most of the $M_\star-\sar$ plane, as shown in the middle panels of Fig.~\ref{fig:msarplane}. We provide the best fit parameters in Table~\ref{tab:modelB}. However, it is not able to recover the observed number of objects at low masses and high-\sar\ (upper left corner). In the upper panels of Fig.~\ref{fig:msarplane} we show the $M_\star-\sar$ plane for our default model A. This mass dependent SARDF model is able to match the observations also in this region (since the SARDF has a higher break at low mass). The AIC$_c$ ratio between the two models prefers our model A with a relative likelihood of $10^{-68}$, providing strong evidence that this model provides a better description for our data set.

An important constraint on this mass dependence in the SARDF is set by the inclusion of the XLF information in our likelihood function, since our sample does not cover a very wide dynamical range  in luminosity. In fact, if we neglect the XLF and only use the data from our XMM-COSMOS sample, such a mass dependence is not strongly required and also the flattening in the SARDF is less strong, while still present. We give the best fit solution without XLF data as model "no XLF" in Tab.~\ref{tab:modelB}. It also matches the XMM-COSMOS sample in the $M_\star-\sar$ plane well (see lower panels in Fig.~\ref{fig:msarplane}). However, this is achieved by proposing a high space density of objects with low $M_\star$ and low \sar, below the luminosity limit of the XMM-COSMOS sample. Such a high space density of low luminosity AGN violates observations of the XLF from deeper surveys \citep{Miyaji2015,Aird2015} and is thus deemed not physical. This is demonstrated in Fig.~\ref{fig:xlf}, where we show the XLF predicted by our bivariate distribution function models. The "no XLF" model (shown by the green solid line) is consistent with our default model (blue line) and the XLF at $z<0.8$, where XMM-COSMOS still probes a relative wide luminosity range. At higher redshift it clearly overpredicts the XLF.
This motivates the additional inclusion of the XLF information in this work, to find a solution to our data set which is consistent with XLF measurements outside of the luminosity range directly covered and thus constrained by our sample.
On the other hand, the XLF alone is degenerate between $M_\star$ and \sar, and thus does not constrain the SARDF and HGMF, as demonstrated by the identical XLF for models A and B. When combining the XMM-COSMOS data with the XLF, the discussed mass and redshift dependencies are required by the data. 
However, the mass dependence of the SARDF is largely driven by the number of few low mass and high \sar\ objects. Future studies, incorporating both deeper surveys (e.g. CDFS, CDFN) and shallower, larger area surveys (e.g. XMM-XXL, Stripe~82X) will be essential to settle this question.

In addition, we also tested a model where we allowed for redshift evolution
in the slope of the $M_\star$-term $\alpha(z) = \alpha_0 + k_\alpha (z-z_0)$, with $z_0$ fixed to 1.1 (hereafter model C). However, we found the best fit  redshift evolution parameter to be consistent with zero $k_\alpha=-0.017\pm 0.043$, thus the inclusion of this additional parameter is not statistically justified. This is also confirmed by the AIC$_c$ ratio between the two models. Similarly, we found that the addition of another parameter which allows for redshift evolution in the break of the $M_\star$-term, $M_\star^*(z) = M_{\star,0}^* + k_{M_\star^*} (z-z_0)$, is not statistically justified (see Table~\ref{tab:aic}).

We also performed the same test of goodness to justify the redshift dependence in the slope of the SAR-term $k_\gamma$ and found this term to be statistically justified.
We provide the relative AIC$_c$ values and the corresponding relative likelihood for different models, compared to our default model (A) in Table~\ref{tab:aic}.

\end{document}